\documentclass[final,nofootinbib,bibnotes]{revtex4}

\usepackage[latin1]{inputenc}
\usepackage[english]{babel}
\usepackage{graphicx}

\usepackage{amsfonts,amsmath,amssymb,multirow}
\numberwithin{equation}{section}

\newcommand{\moy}[1]{\langle #1 \rangle}
\newcommand{\moyqs}[1]{\langle #1 \rangle_\text{qs}}
\newcommand{\Moy}[1]{\Big\langle #1 \Big\rangle}
\newcommand{\Moyqs}[1]{\Big\langle #1 \Big\rangle_\text{qs}}
\newcommand{\rhoqs}{\rho_\text{qs}}

\newcommand{\A}{\mathcal{A}}
\newcommand{\At}{{\mathcal{B}}}
\newcommand{\F}{\mathcal{F}}
\newcommand{\Lin}{\mathcal{L}}
\newcommand{\wb}[1]{\widetilde{#1}}
\newcommand{\wt}[1]{\widetilde{#1}}

\newcommand{\bc}{\alpha}
\newcommand{\lp}{\left(}
\newcommand{\rp}{\right)}

\newcommand{\modif}[1]{#1}

\begin{document}

\title{Quasi-stationary regime of a branching random walk in presence of an
absorbing wall}

\author{Damien Simon}
\email{damien.simon@lps.ens.fr}
\author{Bernard Derrida}
\email{bernard.derrida@lps.ens.fr}
\affiliation{Laboratoire de
Physique Statistique, \'Ecole Normale Sup\'erieure, 24, rue
Lhomond, 75231 Paris Cedex 05, France, }

%

\date{\today}

\begin{abstract}
A branching random walk in presence of an absorbing wall moving at a constant velocity $v$ undergoes a phase
transition as the velocity $v$ of the wall varies. Below the critical velocity
$v_c$, the population has a non-zero survival probability and when the population survives its size grows exponentially. We investigate the histories of the
population conditioned on having a single survivor at some final time
$T$. We study the quasi-stationary regime for $v<v_c$ when $T$
is large. To do so, one can construct a modified
stochastic process which is equivalent to the original process conditioned on having a single survivor at final time $T$. We then use this construction to show that the
properties of the quasi-stationary regime are universal when $v\to v_c$. We also solve exactly a simple version of the problem, the exponential model, for which the study of the quasi-stationary regime can be reduced to the analysis of a single one-dimensional map.
\end{abstract}

\maketitle


\section{Introduction}

Branching random walks are often used as simple models of evolving populations, with or
without selection. They can describe how a population invades an empty
domain, how a favorable mutation spreads
\cite{fisher,hallatschek} or how the fitness of a population evolves under selection \cite{dubertret,kloster}. Recently very simple models \cite{kloster,snyder,brunetgenea,brunetderridacoal} of neo-darwinian evolution \cite{pelitilecture} have been studied, motivated in particular by in vitro experiments \cite{dubertret}. In these models each individual of a population of fixed size is characterized by a single number, \modif{a trait} which we can call its fitness or its adaptability. From one generation to the next, \modif{this trait} is transmitted to the offspring up to small variations which represent the effect of mutations. Then at each generation, only the $N$ fittest individuals survive. It was shown in \cite{brunetderridacoal} that these models can be mathematically reduced to the study of travelling waves in presence of noise, a subject widely studied \cite{doeringmueller,muellersowers,muellermytnik,escudero,moro,pecheniklevine,panja} which appears in many other contexts such as reaction-diffusion problems, disordered systems \cite{brunetdisordered} or QCD \cite{iancumuellermunier,marquetpeschanski,peschanski}.

Another version of the same model of evolution was also considered in \cite{epl-derridasimon} where the size of the population may vary with time and the selection takes place by eliminating at each generation all the individuals which are below some threshold along the fitness axis. When this threshold increases linearly with time, the model reduces to the problem of a branching random walk in presence of an absorbing wall moving at a constant velocity $v$. Depending on how fast the wall moves, the population can either survive or get extinct. The survival probability, in the long time limit, decreases as the velocity $v$ of the wall increases and, above some critical velocity $v_c$, the population gets extinct with probability $1$.

\modif{A model fully equivalent to the branching random walk in presence of an absorbing moving wall, that we study here, is the case of a branching random walk with a drift and a fixed absorbing threshold, which has been introduced in \cite{antal} in order to describe the aging process in a cell proliferation model. In this case too, there is a phase transition to an absorbing state when the drift increases.}

Similar phase transitions to an absorbing state (the empty population) have been extensively studied in the theory of non-equilibrium systems \cite{hinrichsen,odor}, in problems like directed percolation \cite{domanykinzel} or reaction-diffusion processes \cite{redner}. Beyond the characterization of the transition to this absorbing state, there has been an increasing interest to study the process conditioned on the survival at a very late time \cite{dickmanvigidal,dickmanoliveira,harrisharris}.

In many cases, evolutions conditioned to avoid the absorbing empty state lead to a quasi-stationary regime~: for example, in a birth-death process (see \cite{harris62,kestenneyspitzer,khalilifrancon,cattiauxcollet}  and section \ref{sec:toy} below), if one conditions on a given non-zero size at a final time $T$ and looks at the population at a time $t$ such that both $t$ and $T-t$ are much larger than the characteristic relaxation time of the system, the statistical properties depend neither on the initial condition nor on $t$ or $T$ and one observes a quasi-stationary state\footnote{\modif{In the present paper, we use the term ``quasi-stationary'' to describe a situation at time $t$ conditioned on the fact that there is a fixed non-zero number of individuals at a much larger time $T$. Note that ``quasi-stationary'' is often used to describe a different situation where the two times $t$ and $T$ coincide \cite{ferrari95,dickmanvigidal,yaglom}. What we call here a quasi-stationary process  is sometimes called a ``Q-process'' \cite{cattiauxcollet}.}}.
 The probability of the events which contribute to these quasi-stationary regimes decreases with $T$, and therefore they are difficult to observe in simulations. There are however a number of cases for which one can modify the process (i.e. determine the right biais) \cite{hardyharris,harrisharris} to generate the quasi-stationary regime directly.

\begin{figure}
 \begin{center}
 \includegraphics[height=5cm]{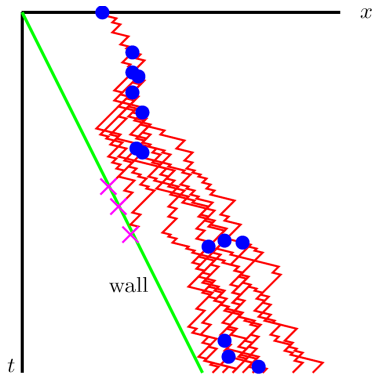}%
\end{center}
\caption{\label{fig1:schema} Branching random walk in presence of an absorbing
wall. The circles
represent branching events and the crosses absorption by the wall.}
\end{figure}

In the present work, we consider the problem of a branching random walk in presence of an absorbing boundary which moves at a constant velocity $v$ (see fig. \ref{fig1:schema} and \cite{epl-derridasimon}). On the infinite line, i.e. without any boundary, the population spreads linearly in time \cite{bramson1,mckean} at a velocity $v_c$ and the number of walkers grows exponentially with time. The presence of a moving absorbing wall introduces a competition between the growth and the extinction due to the wall. When $v>v_c$, the population gets extinct with probability $1$, whereas when $v<v_c$, this extinction probability is strictly less than $1$ and there is a non-zero probability that the population grows exponentially with time. Here, we mostly study the quasi-stationary regime, conditioned on the survival of a few individuals at a very late time $T$, for $v<v_c$. This quasi-stationary regime has already been studied in the mathematical literature, where a modified process was invented \cite{harrisharris,hardyharris} which allows one to generate typical evolutions in this quasi-stationary regime. Our main goal in the present paper is to show how the properties (average size of the population, density profiles) can be calculated in the quasi-stationary regime and how universal expressions for these properties emerge when $v\to v_c$.

The paper is organized as follows~: in section \ref{sec:toy}, we recall how the quasi-stationary regime can be calculated in the case of a simple birth-death process, and how the distribution of the population size becomes universal near the transition.
In section \ref{sec:conditionned}, we extend the approach of section \ref{sec:toy} to the branching random walk process in presence of an absorbing wall.
In section \ref{sec:modified}, we show how the quasi-stationary regime can be generated by a modified process, as already described in \cite{harrisharris,hardyharris} and we obtain explicit expressions of the density in this regime.
In section \ref{sec:universality}, we show that the average size of the population and the average density profile become universal as $v\to v_c$.
In section \ref{sec:exponential}, we analyze a simpler model, the exponential model, for which the whole distribution of the population size can be calculated.

\section{An example of birth-death processes~: the Galton-Watson
process}
\label{sec:toy}

In this section we discuss the simple case of the
Galton-Watson process \cite{harris62,kestenneyspitzer}. Our goal is to explain in this well-known example the
approach that we will use in a more general context in sections
\ref{sec:conditionned}, \ref{sec:modified}, \ref{sec:universality}.

\subsection{Discrete time}

We first consider the discrete time case. The
population is fully characterized by its size $N_t$ at time $t$. At every time
step, each
individual is replaced by $k$ offspring with probability $p_k$ (in particular,
$p_0$ is the probability that an individual leaves no descendance). Let us
define the
extinction probability $Q_e(t)$ as
the probability that $N_t=0$, given that we start with a
single initial individual at time $0$. During the first time step, the initial
individual may branch into $k$ offspring~:
thus,
extinction of the initial lineage after $t+1$ time steps is equivalent to the
extinction of the lineages of its $k$ offspring after $t$ time steps. Since in
this model the descendants of these offspring are independent,
$Q_e(t)$
satisfies the following exact recursion~:
\begin{equation}
 \label{eq1:Qereceq}
Q_e(t+1) =   F\left( Q_e(t) \right)
\end{equation}
where
\begin{equation}
F(Q)=\sum_{k=0}^\infty p_k Q^k
\end{equation}
with the initial condition $Q_e(0)=0$. For large $t$, $Q_e(t)$
converges to a limit $Q_e^*$ and $1-Q_e^*$ is the probability of eternal
survival of the population. This limit $Q_e^*$
is the attractive fixed point of the map $F$, satisfying  $Q_e^* = F( Q_e^*)$. 
The function $F(Q)$ is convex \modif{and satisfy} $F(1)=1$
and $F(0)=p_0>0$. The average number of offspring of an individual is given by
$F'(1)=\sum_k p_k k = \overline{k}$. If $\overline{k}<1$ the stable fixed
point is
$Q_e^* = 1$ and the population dies with probability $1$. On the other hand when
$\overline{k}>1$,
there is a non-zero survival probability $1-Q_e^*$ in the $t\to\infty$ limit.

If one starts at $t=0$ with a
single individual, the distribution
of the population size $N_t$  can be characterized by its generating function
\begin{equation}\label{eq1:defG1}
G_1(t ; \mu) = \moy{ e^{-\mu N_t}}.
\end{equation}
The size $N_t$ is the sum of the contributions of all the
offspring at $t=1$. The independence of the
lineages of the $k$ offspring implies that $G_1$ evolves according to the same equation (\ref{eq1:Qereceq}) as $Q_e(t)$~:
\begin{equation}
\label{eq1:evolG1}
 G_1(t+1 ; \mu) = F\left( G_1(t ;\mu)\right),
\end{equation}
the only difference being in the initial condition $G_1(0 ; \mu)=e^{-\mu}$. Note that the extinction probability is given by the particular case $Q_e(t) = G_1(t ; \infty)$.

\begin{figure}
 \includegraphics[width=0.6\textwidth]{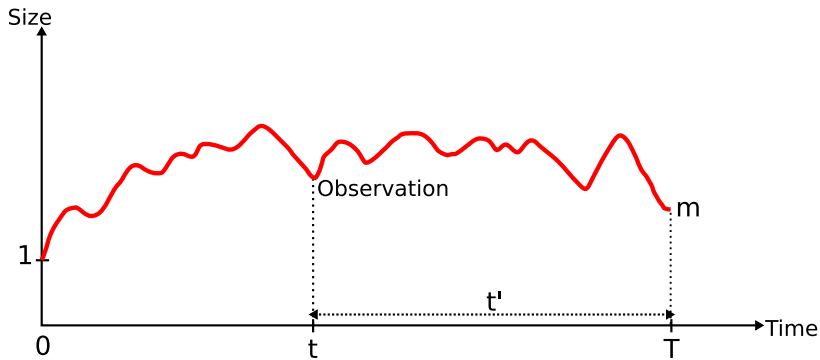}
\caption{\label{fig:timeintervals}\modif{Size $N_t$ of the population at an intermediate time $t$ knowing that the final size is $N_T=m$ at time $T=t+t'$}.}
\end{figure}

For large $T$ \modif{and a given finite size $m$}, the events $N_T=m$
are more and more rare since the population size either vanishes or grows
exponentially. For such events, however, one may be interested in the sizes
$N_t$, $0\leq t \leq T$, conditioned on the fact that $N_T=m$ (as in figure
\ref{fig:timeintervals}). To do so, let us divide the time interval $[0,T]$
into two intervals of lengths $t$ and $t'=T-t$ and consider the two-time
generating functions $G_2(t,t';\mu,\nu)$ defined as~:
\begin{equation}
\label{eq1:defG2}
G_2(t , t'; \mu,\nu) = \moy{ e^{-\mu N_t-\nu N_{t+t'}}}.
\end{equation}
As for $Q_e$ and $G_1$, one can easily show, by considering the first time
step, that $G_2(t , t'; \mu,\nu)$ as a
function of $t$ also evolves according to (\ref{eq1:Qereceq})~:
\begin{equation}
\label{eq1:evolG2}
 G_2( t+1, t' ; \mu,\nu) = F( G_2( t, t' ; \mu,\nu) ),
\end{equation}
with the initial condition
\begin{equation}
G_2( 0, t' ; \mu,\nu) = e^{-\mu} G_1(t';\nu). \label{eq1:G2ini}
\end{equation}
If one expands $G_2( t, t' ; \mu,\nu)$ in powers of $e^{-\nu}$~:
\begin{equation}
\label{eq1:G2decompR}
 G_2(t , t'; \mu,\nu) = \sum_{m=0}^\infty e^{-\nu m}
R_m(t,t';\mu)
\end{equation}
the coefficients $R_m(t,t';\mu)$ are related to the generating functions of
$N_t$ conditioned on the survival of $m$ individuals at time $T$ (i.e.
$N_T=m$)~:
\begin{equation}
\label{eq1:G2decomp_defR}
R_m(t,t';\mu)=  P_m(t+t') \moy{ e^{-\mu N_t} | \text{size $N_{t+t'}=m$}}
\end{equation}
where $P_m(t+t')$ is the probability that $N_{t+t'}=m$. As in the special
case $\mu=0$
\begin{equation}
\label{eq1:G2decompP}
 G_2(t , t'; 0,\nu) = G_1(t+t';\nu) = \sum_{m=0}^\infty e^{-\nu m} P_m(t+t')
\end{equation}
one can extract the probability $P_m(t+t')$ from the expansion (\ref{eq1:G2decompP})
of $G_1(t+t',\nu)$ in order to normalize
$R_m(t,t';\mu)$ in (\ref{eq1:G2decomp_defR}) and get the conditional probability $\moy{ e^{-\mu
N_t} | \text{size $m$ at $t+t'$ }}$.

We are now going to see that $\moy{
e^{-\mu N_t} | \text{size $m$ at $t+t'$ }}$ has a limit as $t$ and
$t'\to\infty$ and to show how to compute the distribution of $N_t$ in this
quasi-stationary regime. Since $G_1$ and $G_2$ evolve according to
(\ref{eq1:evolG1},\ref{eq1:evolG2}),
their long time
behaviours can be extracted from the properties of $F$ near the fixed point
$Q_e^*$. If $u_t$ is a sequence obtained by iterating a map $F$
\begin{equation}
  u_{t+1} = F( u_t) 
\end{equation}
with an initial condition $u_0$. If it converges exponentially to a fixed point $Q_e^*$, one can
write~:
\begin{equation}
\label{eq1:defA}
 u_t \simeq Q_e^* + \Lambda^t\A(u_0)  + o(\Lambda^t), \quad\text{with } \Lambda
=
F'(Q_e^*)
\end{equation}
where the amplitude $\A(u_0)$ is defined by~:
\begin{equation}
\label{eq1:defA2}
 \A(u_0) = \lim_{t\to\infty} \frac{ u_t - Q_e^*}{\Lambda^t}.
\end{equation}
Using (\ref{eq1:defA}) for $G_2$ and $G_1$ together with the
initial
condition
(\ref{eq1:G2ini}) for large $t$ and $t'$ gives~:
\begin{eqnarray}
G_2  (t,t';\mu,\nu) &=& Q_e^* + \Lambda^t \A ( e^{-\mu} G_1(t' ;\nu) ) +
o(\Lambda^t) \nonumber \\
&=& Q_e^* + \Lambda^t \A\left( e^{-\mu} Q_e^* + e^{-\mu}\Lambda^{t'}
\A(e^{-\nu}) +\ldots \right)  + \ldots \label{eq1:G2large} \\
&=& Q_e^* + \Lambda^t \A\left( e^{-\mu} Q_e^* \right)  + \Lambda^{t+t'}
\A'\left(e^{-\mu}Q_e^*\right) e^{-\mu} \A(e^{-\nu}) +\ldots \nonumber
\end{eqnarray}

If one expands $\A(e^{-\nu})=\sum_{m} A_m
e^{-\nu m}$, one gets from (\ref{eq1:G2decompR},
\ref{eq1:G2decompP},\ref{eq1:G2large}) at
leading order in $\Lambda^{t+t'}$ for $m\geq 1$~:
\begin{subequations}
 \label{eq1:PmRm:longtime}
\begin{eqnarray}
 P_m(t+t') &\simeq& \Lambda^{t+t'} \A'(Q_e^*) A_m= \Lambda^{t+t'} A_m,\\
R_m(t,t';\mu) &\simeq& \Lambda^{t+t'} \A'(e^{-\mu}Q_e^*) e^{-\mu} A_m.
\end{eqnarray}
\end{subequations}
From these equations and (\ref{eq1:G2decomp_defR}), it is easy to see that for
$t,t'\to\infty$~:
\begin{equation*}
\moy{ e^{-\mu N_t} | \text{size $N_{t+t'}=m$ }}  \xrightarrow{t,t'\to\infty} \moyqs { e^{-\mu
N} }
\end{equation*}
with \begin{equation}
\label{eq1:quasistat-A}
 \moyqs { e^{-\mu N} } = \A'(e^{-\mu}Q_e^*) e^{-\mu} .
\end{equation}
This shows that the knowledge of the amplitude $\A$ defined in (\ref{eq1:defA2})
determines the whole distribution of the population size in the quasi-stationary
regime \cite{khalilifrancon}. 

In general there is
not a closed expression of $\A$ for a given map $F$. One can however relate the
expansions of $\A$ and of $F$ in the
neighbourhood of the fixed point $Q_e^*$. The definition (\ref{eq1:defA2}) implies
that $\A$ satisfies~:
\begin{eqnarray}
 \Lambda \A(Q)  &=& \A( F(Q) ), \label{eq1:relationA-F} \\
  \A(Q_e^*)&=& 0, \quad \A'(Q_e^*)= 1 . \nonumber 
\end{eqnarray}
By taking successive derivatives of (\ref{eq1:relationA-F}) at $x=Q_e^*$, one can relate the Taylor expansions of $\A$ and $F$
\begin{eqnarray*}
 \A( Q_e^* + \epsilon) = \epsilon + \frac{\epsilon^2}{2!} \A^{(2)} +
\frac{\epsilon^3}{3!} \A^{(3)} + \ldots \\
F( Q_e^* + \epsilon) = Q_e^* + \Lambda
\epsilon + \frac{\epsilon^2}{2!}
F^{(2)} +
\frac{\epsilon^3}{3!} F^{(3)} + \ldots
\end{eqnarray*}
and get~:
\begin{subequations}
 \label{eq1:A-F-discrete}
\begin{eqnarray}
  \A^{(2)} &=& \frac{ F^{(2)} }{\Lambda-\Lambda^2} \\
  \A^{(3)} &=& \frac{ F^{(3)}}{\Lambda(1-\Lambda)(1+\Lambda)}  + \frac{3
 (F^{(2)})^2}{\Lambda ( 1+\Lambda)( 1-\Lambda)^2 }
\end{eqnarray}
\end{subequations}
etc. Then, the moments of $N_t$ in the quasi-stationary regime can be
obtained by taking successive derivatives of (\ref{eq1:quasistat-A}) at
$\mu=0$~:
\begin{subequations}
\label{eq1:momentsn-A}
\begin{eqnarray}
 \moyqs{ N } &=& 1 + Q_e^*\A^{(2)}  \\
 \moyqs{ N^2 } &=& 1 + 3 Q_e^* \A^{(2)}  + (Q_e^*)^2\A^{(3)},
\text{etc.}
\end{eqnarray}
\end{subequations}
As one approaches the transition point (i.e. $\overline{k}\to 1$
and $Q_e^*\to 1$), expressions (\ref{eq1:A-F-discrete},\ref{eq1:momentsn-A})
diverge since $\Lambda\to 1$
and the moments scale as~:
\begin{eqnarray}
 \moyqs{ N } &\simeq & \A^{(2)} \simeq \frac{F^{(2)} }{1-\Lambda} \\
\frac{ \moyqs{ N^2 }}{ \moyqs{ N }^2} &\simeq&
\frac{\A^{(3)}}{(\A^{(2)})^2}\simeq \frac{3 \Lambda }{(1+\Lambda)} +
\frac{ (1-\Lambda)F^{(3)}}{  (1+\Lambda) \big(F^{(2)}\big)^2}
\end{eqnarray}
and so on for higher moments $\moyqs{N^p}/\moyqs{N}^p$, $p>2$.
When the first moments of the number of offspring are finite (for example when
$\overline{k^3}, \overline{k^4}<\infty$), the ratios such as ${ (1-\Lambda)F^{(3)}}/({
(1+\Lambda) \big(F^{(2)}\big)^2}) $
go to zero as $\Lambda \to 1$ and  one gets~:
\begin{subequations}
\label{eq1:discrete:N:univ}
\begin{eqnarray}
 \moyqs{N} & \simeq &  \frac{F^{(2)}(Q_e^*) }{1-F'(Q_e^*)} \\
 \frac{ \moyqs{N^2} }{\moyqs{N}^2 } &\to & \frac{3}{2} \\
 \frac{ \moyqs{N^3} }{\moyqs{N}^3 } &\to & 3, \quad\text{etc.}
\end{eqnarray}
\end{subequations}
One can notice that these expressions are valid on both sides of the
transition, i.e. for $\overline{k}\to
1^+$ and $\overline{k}\to
1^-$ but do not depend otherwise on the $p_k$'s. This agrees with the fact (see \cite{harris62} and section
\ref{subsec:toy:univ} below) that the distribution of $N_t$ in the quasi-stationary
regime becomes universal near the transition $\overline{k}=1$.

\subsection{Continuous time version}
\label{sec:toy-continuous}

We now consider the continuous time version of the previous birth-death
process~: at any
instant $t$, the population is fully
characterized by its size $N_t$. During every infinitesimal time $dt$, each
individual either dies with probability $\beta_0 dt$ or branches into $k$
offspring with probability $\beta_k dt$. If one divides the time interval
$[0,t+dt]$ into the first interval $dt$ and the remaining interval
$[dt,t+dt]$, the independence of the
evolution of the offspring implies that the extinction probability $Q_e(t)$
satisfies as in (\ref{eq1:Qereceq})~:
\begin{equation}
 Q_e(t+dt) = \left(1-\sum_{k\geq 0} \beta_k dt\right) Q_e(t) + \sum_{k\geq 0}
\beta_k dt Q_e(t)^k
\end{equation}
which gives in the limit $dt\to 0$~:
\begin{equation}
 \partial_t Q_e(t) = F(Q_e(t) ),\quad F(Q) = \sum_{k\geq 0} \beta_k (Q^k-Q).
\end{equation}
This is the continuous time version of (\ref{eq1:Qereceq}). In the
long time limit, $Q_e(t)$ converges to the stable fixed
point $Q_e^*$ of $F$~:
\begin{equation}
 F( Q_e^*) = 0, \quad F'(Q_e^*)=\lambda < 0.
\end{equation}
The transition occurs when $Q_e^*\to 1$ and $\lambda \to 0$. It corresponds to a
growth rate
\begin{equation}
 \alpha=
\sum_{k\geq 0}\beta_k (k-1)
\end{equation}
which vanishes. If $\alpha <0$, $Q_e^*=1$ whereas
for $\alpha >0$, one has $Q_e^* <1 $.

The generating functions $G_1$ and $G_2$ can be
introduced as in (\ref{eq1:defG1},\ref{eq1:defG2}) and the discussion of
their long time behaviours is the same as in (\ref{eq1:G2large}). For a flow
$u(t)$ such that $\partial_t u(t) =
F(u(t))$ and $u(0)=u_0$, one can define an amplitude $\A$ as in
(\ref{eq1:defA}) by~:
\begin{equation}
 \A(u_0) =  \lim_{t\to\infty} \frac{ u(t) - Q_e^*}{e^{\lambda t}} .
\end{equation}
The generating functions of the size $N_t$ in the quasi-stationary regime are
still given by (\ref{eq1:quasistat-A}) but the definition of $\A$ for continuous
time
leads to the following continuous time version of (\ref{eq1:relationA-F})~:
\begin{equation}
\label{eq1:defAcontinuous}
 \lambda \A(Q) = \A'(Q) F(Q) .
\end{equation}
As in (\ref{eq1:A-F-discrete}), the expansions of $\A$ and $F$ near the fixed point $Q_e^*$ can be related
through~:
\begin{subequations}
 \label{eq1:A-F:continuous:exp}
\begin{eqnarray}
 \A^{(2)} &=& -\frac{F^{(2)}}{\lambda}, \\
 \A^{(3)} &=& \frac{3}{2} \left( \frac{ F^{(2)}}{\lambda}\right)^2 -
\frac{F^{(3)}}{2\lambda} .
\end{eqnarray}
\end{subequations}
The average size is given by $\moyqs{N}= -F^{(2)}/\lambda$ and diverges as one
approaches the transition $\alpha\to 0$. For models with finite moments
$\overline{k^n}$, the first moments $\moyqs{N^n}$ obtained from the expansion
(\ref{eq1:A-F:continuous:exp}) of $\A$ go to the same values as in
(\ref{eq1:discrete:N:univ}).

As (\ref{eq1:defAcontinuous}) is a first order
differential equation, one can fully determine $\A$ using
$\A'(Q_e^*)=1$ (which follows from definition (\ref{eq1:defAcontinuous}))~:
\begin{equation}
\label{eq1:A-intrep}
 \A(Q) = (Q-Q_e^*) \exp\left[  \int_{Q_e^*}^Q \left(
\frac{\lambda}{F(y)}-\frac{1}{y-Q_e^*}\right) dy \right].
\end{equation}

\subsection{Universality of the stationary regime}
\label{subsec:toy:univ}

Near the transition, $Q_e^*$ is close to $1$, $\lambda$ is small and for $Q$
close to $Q_e^*$, one can approximate $F(Q)$ by the first two terms of its
quadratic expansion near $Q_e^*$ (when $\overline{k^2} <\infty$)~:
\begin{equation}
\label{eq1:FexpansionQestar}
 F(Q) \simeq \lambda (Q-Q_e^*) + B (Q-Q_e^*)^2.
\end{equation}
Then (\ref{eq1:A-intrep}) becomes near $Q_e^*$~:
\begin{equation}
 \A (Q) \simeq \frac{\lambda (Q -Q_e^*)}{ \lambda + B (Q-Q_e^*)}
\end{equation}
and from (\ref{eq1:quasistat-A}), one gets that for $\mu$ small~:
\begin{equation}
\label{eq1:N:univ}
 \moyqs{ e^{-\mu N} } \simeq \frac{1}{ \left( 1-\frac{B}{\lambda}\mu\right)^2}.
\end{equation}
Close to the transition, $B\simeq F''(1)/2\simeq
(\overline{k^2}-\overline{k})/2$ and this shows that in
the quasi-stationary regime~:
\begin{equation}
 \moyqs{N} \simeq  -\frac{\overline{k^2}-\overline{k}}{\lambda} 
\end{equation}
and the distribution of the ratio $x=N/\moyqs{N}$ becomes universal
\cite{harris62}~:
\begin{equation}
 P(x) \simeq 4x e^{-2x}
\end{equation}
with the ratios of the first moments given by (\ref{eq1:discrete:N:univ}).

The case of slowly decreasing branching rates $\beta_k$ for which the second
moment $\overline{k^2}$ is infinite leads to other universal distributions near
the transition, as shown in \cite{slack} and in appendix
\ref{appendix:heavytail}.


\section{Quasi-stationary regime for a branching random walk in presence of an
absorbing wall}
\label{sec:conditionned}

\subsection{Evolution of the extinction probability}

We now consider a population for which each individual is characterized by its
position $x$ relative to an absorbing wall. Each
individual diffuses and branches into $k$ offspring with probability $\beta_k
dt$ during $dt$ (for simplicity we set $\beta_0=0$, \emph{i.e.} there is no spontaneous death ; also there is no dependence on $\beta_1$ as it has no effect on the population). The absorbing wall
moves at constant velocity $v$ and every individual crossing the
wall disappears instantaneously (see figure \ref{fig1:schema}). In absence of the wall, the population grows
exponentially and occupies a region which spreads  in space linearly at a constant velocity $v_c$ \cite{bramson1,mckean}. The presence of the wall
introduces a competition between the exponential growth and the absorption by
the wall. For $v>v_c$, the wall moves faster than the population spreads \cite{epl-derridasimon} and
the population dies with probability $1$.

If one starts at time $t=0$ with a single
individual at distance $x$ from the moving wall, \modif{the extinction probability
$Q_e(x,t)$ (\emph{i.e.} the probability that all the descendants have been absorbed by the wall during the interval $[0,t]$) evolves according to~:}
\begin{equation}
\begin{split}
Q_e(x+vdt, t+dt ) =& \int \frac{e^{-\eta^2/4}}{\sqrt{4\pi}}
Q_e(x+\eta\sqrt{dt}, t) d\eta 
 \\
&+ \sum_{k=2}^\infty \beta_k
\left(Q_e(x,t)^k-Q_e(x,t)\right) dt  .\label{eq2:Qe_pde_deriv}
\end{split}
\end{equation}
\modif{
To derive this equation, we
divide, as in section \ref{sec:toy}, the time interval $[0,t+dt]$ into a first
time step $dt$ and a second
interval $[dt,t+dt]$. During
the first infinitesimal time step $dt$, the individual
diffuses (its position is shifted by an amount $\eta \sqrt{dt}$ where $\eta$ is
a Gaussian random variable such that $\moy{\eta}=0$ and $\moy{\eta^2}=2$)
and may branch into $k\geq 2$ individuals with rate $\beta_k$. If this branching event happens, then, during the
second time interval $[dt,t+dt]$, all the $k$ offspring present at time $dt$ have independent histories.}

In the limit $dt\to 0$, this becomes ~:
\begin{equation}
 \label{eq2:Qe_pde}
 \partial_t Q_e = \F( Q_e)
\end{equation}
where the functional $\F$ is defined as~:
\begin{eqnarray}
 \label{eq2:deffunctionalF}
\F( Q ) &=& \partial_x^2 Q(x) - v\partial_x Q(x)+ g( Q(x) ) \\
g\left(q\right) &=&  \sum_{k=2}^\infty \beta_k \Big( q^k -q\Big).
\label{eq2:defnonling}
\end{eqnarray}

$Q_e$ evolves according to a travelling wave equation \cite{vansaarloos}. Here however the  evolution is limited to a semi-infinite line with the boundary condition $Q_e(0,t)=1$ on the wall. The non-linear function $g(q)$ contains all the information on the branching rates $\beta_k$ of the walk.
In the long time limit $t\to\infty$, the extinction probability $Q_e(x,t)$
converges to the stable solution $Q_e^*$ of $\F(Q_e^*)=0$ and therefore
satisfies~:
\begin{equation}
\label{eq:defQestar}
 \partial_x^2 Q_e^* - v \partial_x Q_e^* + g\left( Q_e^*(x)\right) = 0.
\end{equation}

For $v>v_c$, the wall moves faster than the spreading velocity $v_c$ of
the population, so that all the population gets absorbed and for large $t$, $Q_e(x,t)$ has the shape of a front moving at velocity $v-v_c$. Therefore $Q_e(x,t)$ converges to
the unique stable fixed point $Q_e^*=1$ (the population becomes
extinct with probability $1$ whatever the initial distance $x$ to the wall is).

For $v<v_c$, there exists a non
trivial fixed point $Q_e^*$ with $Q_e^*(0)=1$ and $Q_e^*(x)\to 0$ for
$x\to\infty$ and $Q_e(x,t)\to Q_e^*(x)$ as $t\to\infty$, with relaxation times
$\tau_n$~:
\begin{equation}
 Q_e(x,t)\simeq Q_e^* +  A_1 \phi_1(x)e^{-t/\tau_1}+ \ldots 
\end{equation}
These relaxation times $\tau_n$ are related through $\lambda_n=-1/\tau_n$ to the eigenvalues $\lambda_n$ of the
linear
operator $\Lin$ obtained by linearizing the functional $\F(Q_e^*+\epsilon
\phi)=\epsilon\Lin[\phi] +o(\epsilon) $ around the fixed point $Q_e^*$. The
eigenvectors $\phi_n(x)$ and the eigenvalues $\lambda_n$ satisfy~:
\begin{equation}
 \Lin[\phi_n]=\lambda_n \phi_n
\end{equation}
with the linear operator $\Lin$ given by~:
\begin{equation}
\label{eq2:Qe:linop:def}
 \Lin[\phi] = \partial_x^2 \phi - v \partial_x \phi + g'\left( Q_e^*(x) \right) \phi
\end{equation}
Defining $\psi(x) = \phi(x) e^{-vx /2}$ maps (\ref{eq2:Qe:linop:def}) to a Schrödinger
equation
\begin{equation}
\label{eq:schrodinger}
 -\partial_x^2\psi(x) + V(x) \psi(x) = (-\lambda) \psi(x)
\end{equation}
with a potential $V$ given by~:
\begin{equation}
\label{eq:schrodinger:potential}
 V(x) = \frac{v^2}{4}  - \sum_k \beta_k (kQ_e^*(x)^{k-1} -1)
\end{equation}
(note that for $v> v_c$, one has $Q_e^*(x)=1$, so the potential $V$ is constant and the spectrum is continuous. For $v>v_c$, $Q_e(x,t)$ converges  to $Q_e^*(x)$
in a non uniform way, so that there are always some $x$ for which
$Q_e(x,t)-Q_e^*(x)$ is not small and the long time behaviour of $Q_e(x,t)$ cannot be understood by linearizing around the fixed point $Q_e^*(x)=1$).

\modif{As $v$ approaches $v_c$, the region where $Q_e^*(x)$ is close to $1$ increases (and diverges when $v=v_c$). As a consequence, as $v\to v_c$, the potential well $V(x)$ in (\ref{eq:schrodinger:potential}) has a finite depth but an increasing length and this gives rise to  a discrete spectrum for small eigenvalues}. We will calculate the eigenvalues $\lambda_n$ of $\Lin$ in this limit $v\to v_c$ in section \ref{sec:eigenvalues} and appendix \ref{appendix:eigenvalues}.

In the rest of this paper we will only consider the case $v<v_c$ when the
smallest
eigenvalues of $\Lin$ form a discrete spectrum (only the construction of section
\ref{subsec:modifiedprocess} will be also valid for $v>v_c$ as one does not
need there the slowest eigenvalue to be isolated).

In constrast to
(\ref{eq1:defG1}), the population is now not only characterized by its size
$N_t$
but also by the positions of all the individuals. One can however follow the
same approach as in section \ref{sec:toy}. Let us
introduce the generating function $G_1$~:
\begin{equation}
\label{eq:def:genefuncG1}
 G_1(x,t; f) = \Moy{ \prod_{i=1}^{N_t} e^{-f(x_i^{(t)})} }
\end{equation}
where $f$ is now a positive test function of $x$ and plays the role of $\mu$ in
(\ref{eq1:defG1}), $N_t$ is the size of the population
at time $t$ generated by an initial individual at $x$ and the $x_i^{(t)}$
are the positions relatively to the wall of all its offspring at
$t$. The independence of the lineages implies that $G_1$ for the initial
individual is the product of the $G_1$'s of its offspring after the
first time step $dt$. Therefore, the evolution of $G_1(x,t;f)$ can be derived as
in (\ref{eq2:Qe_pde_deriv}) for $Q_e(x,t)$ and one gets~:
\begin{equation}
 \label{eq2:g1equadiff}
\partial_t G_1(x,t ; f)  = \F\big(G_1(x,t;f)\big).
\end{equation}
Absorption by the wall implies $G_1(0,t;f)=1$ for all $t>0$ and, for a single initial
individual at $x$ when $t=0$, the initial condition for $G_1$ is~:
\begin{equation}
\label{eq2:iniG1}
 G_1(x,0 ;f) = e^{-f(x)}.
\end{equation}

\subsection{Conditioning on survival}

As in section \ref{sec:toy}, one can then try to calculate the size $N_t$ of
the population or the density $\rho(x)$ at time $t$ conditioned on the
survival of $m$ individuals at later time $T=t+t'$ (see fig.~\ref{fig:timeintervals}). To do so we introduce
the two-time generating function at
$G_2(x, t, t'; \mu,\nu)$  defined as in (\ref{eq1:defG2})~:
\begin{equation}
\label{eq:defgeneG2}
G_2(x, t, t'; f,\nu) = \Moy{ \exp\Big( -\sum_{i=1}^{N_t} f( x_i^{(t)}) - \nu N_{t+t'} \Big)}
\end{equation}
given that at $t=0$ there is a single individual at distance $x$ from the wall.
Once more, one can show that $G_2(x,t,t'; f,\nu)$ \modif{(as a function of $x$ and $t$ only)} evolves as $G_1(x,t;f)$, i.e. according to (\ref{eq2:g1equadiff})  with the
following initial
condition~:
\begin{equation}
\label{eq2:iniG2}
 G_2(x,0,t'; f,\nu) = e^{-f(x)} G_1(x,t';\nu).
\end{equation}
As in (\ref{eq1:G2decompR},\ref{eq1:G2decomp_defR},\ref{eq1:G2decompP}), one can expand $G_2$ in powers of $e^{-\nu}$~:
\begin{eqnarray}
\label{eq2:g2expansionR}
 G_2(x, t, t'; f, \nu) &=& \sum_{m=0}^\infty e^{-\nu m} R_m(x,t,t'; f) \\
 G_2(x, t, t'; 0, \nu) &=& G_1(x,t+t';\nu)  = \sum_{m=0}^\infty e^{-\nu m}
P_m(x,t+t')  \label{eq2:g2expansionP}
\end{eqnarray}
where $P_m(x,t+t')$ is the probability that the initial individual located at
$x$ has exactly $m$ living descendants at $t+t'$ and the generating function of $f$
conditioned on observing a size $m$ at $t+t'$ is given by~:
\begin{equation}
\label{eq2:linkg2conditionned}
 \Moy{ \exp\Big( -\sum_{i=1}^{N_t} f(x_i^{(t)})\Big) \Big| \text{size $m$ at
$t+t'$} } =  \widetilde{R}_m (x, t, t' ; f)
\end{equation}
where $\widetilde{R}_m (x, t, t' ; f)$ is defined as~:
\begin{equation}
 \label{eq2:linkg2conditionned-defRtilde}
\widetilde{R}_m (x, t, t' ; f) =\frac{R_m(x,t,t';f)}{P_m(x,t+t')}
\end{equation}
This shows that all the information about the regime $0<t<T$, conditioned on
having the final size $N_T=m$,  is contained in the functions
$\widetilde{R}_m(x,t,t';f)$. 
Eq. (\ref{eq2:iniG2},\ref{eq2:g2expansionR},\ref{eq2:g2expansionP}) are
the analogues of (\ref{eq1:G2ini},\ref{eq1:G2decompR},\ref{eq1:G2decompP}) in
section \ref{sec:toy}.

 Inserting the
expansions (\ref{eq2:g2expansionR},\ref{eq2:g2expansionP}) into
(\ref{eq2:g1equadiff}) leads to differential equations for the $R_m$'s and the
$P_m$'s. At lowest order, $R_0$ and $P_0$ satisfy (\ref{eq2:g1equadiff})
and, at first order, $R_1$ and $P_1$ are related to $R_0$ and $P_0$
through~:
\begin{subequations}
\begin{eqnarray}
\label{eq2:r1equadiff}
 & &\partial_t R_1 = \partial_x^2 R_1 - v \partial_x R_1 + \sum_{k=2}^\infty
\beta_k (k R_0^{k-1} -1) R_1 \\
\label{eq2:p1equadiff}
& &\partial_t P_1 = \partial_x^2 P_1 - v \partial_x P_1 + \sum_{k=2}^\infty
\beta_k (k P_0^{k-1} -1) P_1 \\
& &R_1(x,0,t';f)=e^{-f(x)}P_1(x,t'), \quad P_1(x,0)=1
\label{eq2:p1r1initial}
\end{eqnarray}
\end{subequations}
By expanding further $G_2$ in powers of $e^{-\nu}$, one could obtain in the same way the evolution equations of the $R_m$'s and $P_m$'s and from (\ref{eq2:iniG2}) their initial values.

\subsection{The quasi-stationary regime}
\label{subsec:dynamical}

\modif{In this subsection, we show how one can} generalize the expression (\ref{eq1:quasistat-A}) of
section \ref{sec:toy} relating the properties of the
quasi-stationary regime to those of the functional $\F$
defined in
(\ref{eq2:deffunctionalF}) near the fixed point $Q_e^*$ for
$v<v_c$. \modif{An alternative (easier) approach to calculate these properties will be described and used in sections \ref{sec:modified} and \ref{sec:universality}}.

The generating functions $G_1$ and $G_2$ evolve according to
(\ref{eq2:g1equadiff})
and, as soon as the test function $f$ is positive, $G_1$ and $G_2$ converge to
the
non-trivial fixed point $Q_e^*$ of $\F$ when $v<v_c$. In order to analyze the
long time behaviour of $G_1$ and $G_2$, one can expand $\F$ around $Q_e^*$~:
\begin{equation}
 \label{eq:Fexpansion}
 \F(Q_e^*+ \epsilon\phi) = \F(Q_e^*) + \epsilon \Lin[\phi] +
\frac{\epsilon^2}{2!} \F^{(2)}[\phi,\phi] +\ldots
\end{equation}
where the $n$-linear symmetrical functionals $\F^{(n)}$ are defined for any
arbitrary set of $n$ functions $\psi_i$ as~:
\begin{equation}
\label{eq:Fexpansion-def}
 \F^{(n)}[\psi_1,\ldots,\psi_n] = \frac{d^n \F( Q_e^* +
s_1\psi_1+\ldots +s_n\psi_n)   }{ds_1\ldots ds_n}  \Big|_{\mathbf{s}=0}.
\end{equation}

As discussed after (\ref{eq:schrodinger:potential}), the linear operator
$\Lin$ does not always have a discrete
spectrum (in particular for $v>v_c$). We will assume here that $v<v_c$ and that
at least the smallest eigenvalue $\lambda_1$ is isolated. If the first
eigenvalue is not isolated, all the construction below breaks down
and the quasi-stationary state might not exist \cite{epl-derridasimon}.

As $Q_e$, $G_1$ and $G_2$ all evolve according to (\ref{eq2:g1equadiff}), one
can study the dynamical system~:
\begin{equation}
\label{eq:flow}
 \begin{cases}
  \partial_t u(x,t) &= \F\left( u(x,t) \right)  \\
  u(x,0) &= u_0(x) 
 \end{cases}
\end{equation}
When $u(x,t)\to Q_e^*(x)$ as $t\to\infty$ and if $\Lin$ has a discrete spectrum
for small eigenvalues, the
convergence of $u(x,t)$ is exponential. Then
one can introduce as in (\ref{eq1:defA2}) the amplitude $\A$ of the first eigenvector $\phi_1$ of $\Lin$ (with the
largest
relaxation time $\tau_1$)~:
\begin{equation}
\label{eq2:defAcontinuous}
 u(x,t) = Q_e^* +  \phi_1(x) e^{-t/\tau_1} \A\left( u_0(\cdot) \right)+\ldots
\end{equation}
but now the argument of the amplitude $\A$ is a function (the
initial condition $u_0(x)= u(x,0)$).

From the initial conditions (\ref{eq2:iniG1},\ref{eq2:iniG2}) of $G_1$ and
$G_2$, the generating functions $G_1(x,t;f)$ and $G_2(x,t,t';f,\nu)$ thus have
the following behaviours as $t\to\infty$~:
\begin{eqnarray*}
 G_1(x,t;f) &=& Q_e^*(x) + \phi_1(x)e^{-t/\tau_1} \A( e^{-f}) +
o(e^{-t/\tau_1}); \\
 G_2(x,t,t';f,\nu) &=& Q_e^*(x) + \phi_1(x) e^{-t/\tau_1} \A\left(
e^{-f}G_1(\cdot,t';\nu) \right)+
o(e^{-t/\tau_1}):
\end{eqnarray*}
Thus for large $t$ and $t'$, at leading orders, $G_2(x,t,t';f,\nu)$ becomes~:
\begin{eqnarray}
 G_2(x,t,t';f,\nu) &\simeq& Q_e^*(x) + \phi_1(x) e^{-{t/\tau_1}} \A\left(
e^{-f} \left( Q_e^* + e^{-{t'/\tau_1}}\phi_1 \A( e^{-\nu})\right)
\right) \nonumber \\
&\simeq& Q_e^*(x) + \phi_1(x)e^{-t/\tau_1} \A\left( e^{-f}Q_e^* \right)
  \label{eq2:G2f:large}\\
& &+
\phi_1(x)e^{-{(t+t')/\tau_1}} \frac{d}{ds} \A\left( e^{-f}Q_e^* + s
e^{-f}\phi_1\right) \Big|_{s=0} \A(e^{-\nu}) +\ldots \nonumber
\end{eqnarray}
As $ G_2(x,t,t';0,\nu)=G_1(x,t+t';\nu)$, one can proceed as in (\ref{eq1:PmRm:longtime}) by writing $\A(e^{-\nu}) =
\sum_m A_m e^{-\nu m}$ and get for $R_m$ and $P_m$ defined in
(\ref{eq2:g2expansionR},\ref{eq2:g2expansionP}) for $m\geq 1$~:
\begin{subequations}\label{eq:largeRmPm}
\begin{eqnarray}
  R_m(x,t,t';f) &\simeq& A_m \phi_1(x) e^{-(t+t')/\tau_1}    \frac{d}{ds}\A\left(
e^{-f}Q_e^* + s
e^{-f}\phi_1\right)\Big|_{s=0} \\
  P_m(x,t+t') &\simeq& A_m \phi_1(x) e^{-(t+t')/\tau_1} 
\end{eqnarray}
\end{subequations}
Thus, the generating function of $f$ given by
(\ref{eq2:g2expansionR}), when $t,t'\to\infty$, has a finite limit~:
\begin{equation}
\label{eq:linkA1qs}
\lim_{t,t'\to\infty} \Moy{ \exp\Big( -\sum_{i=1}^{N_t} f(x_i^{(t)})\Big) \Big|
\text{size $N_{t+t'}=m$} } = \Moyqs{ \exp\Big( -\sum_{i=1}^{N_t}
f(x_i^{(t)})\Big) }
\end{equation}
given by~:
\begin{equation}
 \label{eq2:quasistat-A}
 \Moyqs{ \exp\Big( -\sum_{i=1}^{N_t} f(x_i^{(t)})\Big) } =\frac{d}{ds} \A\left(
e^{-f}Q_e^* + s
e^{-f}\phi_1\right)\Big|_{s=0}
\end{equation}
(note that as in (\ref{eq2:defAcontinuous}) the argument of $\A$ is a function).
We see that, as in (\ref{eq1:quasistat-A}), the properties of the quasi-stationary state
are determined in (\ref{eq2:quasistat-A}) by the amplitude $\A$ (expression
(\ref{eq2:quasistat-A}) does not depend
anymore on $t$,
$t'$, nor on the value of $m$ used to condition the population
size at $t+t'$, nor on
the position $x$ of the initial individual).

From the definition (\ref{eq:flow},\ref{eq2:defAcontinuous}), one can see that the amplitude $\A$ satisfies~:
\begin{equation}
\label{eq:A1-F}
 \lambda_1 \A(u) = \frac{d}{d\tau} \A(u+ \tau\F(u))\Big|_{\tau=0}.
\end{equation}
As in section \ref{sec:toy}, this equation can be used to relate $\A$ and $\F$. Note that equations similar
to
(\ref{eq:A1-F}) appear in other contexts, in particular in
the renormalization group theory where $\A$ is called a non-linear
scaling field \cite{wegner}. As in (\ref{eq:Fexpansion}), one can expand $\A$
around
$Q_e^*$~:
\begin{equation}
\label{eq:Aexpansion}
 \A( Q_e^* +\epsilon \phi) = \A(Q_e^*) + \epsilon \A^{(1)}[\phi] +
\frac{\epsilon^2}{2!} \A^{(2)}[\phi,\phi] +  \ldots
\end{equation}
Successive derivatives of (\ref{eq:A1-F}) as defined in
(\ref{eq:Fexpansion-def}) in the directions
of the eigenvectors $\phi_n$ of $\Lin$
allow one to relate the expansions (\ref{eq:Aexpansion},\ref{eq:Fexpansion}) of $\A$ and $\F$~:
\begin{subequations}
 \label{eq:expansionsA-F}
\begin{eqnarray}
\A^{(1)}[\phi_i] &=&  \delta_{i,1} 
\label{eq2:AF1}\\
\A^{(2)}[\phi_i,\phi_j] &=&  \frac{1}{\lambda_1-\lambda_i-\lambda_j}
\A^{(1)}[ \F^{(2)}[\phi_i,\phi_j]]  \label{eq2:AF2}\\
\A^{(3)}[\phi_i, \phi_j,\phi_k] &=&
\frac{1}{\lambda_1-\lambda_i-\lambda_j-\lambda_k} \\
& &\times \Big( \A^{(1)}\big[
\F^{(3)}[\phi_i,\phi_j,\phi_k]\big] + \label{eq2:AF3} \A^{(2)}\big[\F^{(2)}[\phi_i,\phi_j],\phi_k\big] \nonumber \\
& &+ \A^{(2)}\big[
\F^{(2)}[\phi_i,\phi_k],\phi_j\big] + \A^{(2)}\big[
\F^{(2)}[\phi_j,\phi_k],\phi_i\big] \Big) \nonumber
\end{eqnarray}
\end{subequations}
Then, from (\ref{eq2:quasistat-A}), one can obtain the properties of the
quasi-stationary regime. For example, the moments of the number of individuals
in the quasi-stationary
state are obtained by taking $f=\mu$ constant and taking successive derivatives
at $\mu=0$~:
\begin{eqnarray}
 \moyqs{ N } &=&  -\frac{d^2}{dsd\mu} \A( Q_e^* e^{-\mu} + s\phi_1 e^{-\mu})
\Big|_{\mu=s=0}  \nonumber \\ 
&=& \A^{(1)}[\phi_1] + \A^{(2)}[\phi_1, Q_e^*] = 1
+\A^{(2)}[\phi_1, Q_e^*] \label{eq:Nmoy-A}\\
 \moyqs{N^2} &=&  \frac{d^3}{dsd\mu^2} \A( Q_e^* e^{-\mu} + s\phi_1 e^{-\mu})
\Big|_{\mu=s=0} \nonumber\\
&=& \A^{(1)}[\phi_1] + 2 \A^{(2)}[\phi_1, Q_e^*] +
\A^{(3)}[\phi_1, Q_e^*, Q_e^*] \label{eq:Nvar-A}
\end{eqnarray}
These expressions are exact and valid as long as the functional $\F$ has a
non-trivial fixed point $Q_e^*$ and an isolated largest relaxation time
$\tau_1$. In order to go
further, one needs to know more precisely $Q_e^*$ and $\phi_1$. This will be done in section \ref{sec:universality} in the scaling regime $v\to v_c$.

\section{A modified process to describe conditioned histories}
\label{sec:modified}

Before analyzing this scaling regime we are going to show
that one can construct a
modified process reproducing the history $0 \leq t \leq T$ of the branching
random walk of section \ref{sec:conditionned} conditioned on a final
size $N_T=1$. This construction is valid both below and above the
critical velocity $v_c$. The quasi-stationary state for $v<v_c$ discussed in
section \ref{subsec:dynamical} will appear as a particular case and its average
profile will be calculated in section \ref{subsec:modproc:profile}.

\subsection{Construction of the modified process}
\label{subsec:modifiedprocess}

One way of describing branching random walks conditioned on the survival of a single individual is to distinguish the path of the surviving particle from the other ones. In the mathematical literature \cite{harrisharris,hardyharris}, this particle is called the \emph{spine} and has its dynamics modified in order to prevent it from dying. The rest of the population is generated from this special particle. We show in the present section that this construction is general for branching random walks conditioned on having exactly one survivor at finite time $T$ and requires only the knowledge of the extinction probability $Q_e(x,t)$ and the probability $P_1(x,t)$ that there is exactly one survivor at $t$.

To condition the evolution of the population on
the survival of one individual at time $t+t'$, 
equations (\ref{eq2:linkg2conditionned},
\ref{eq2:linkg2conditionned-defRtilde}) show that one should
consider the ratio $\widetilde{R}_1 (x, t,
t' ; f) = {R_1(x,t,t'; f)}/{P_1(x,t+t')}$.
From
(\ref{eq2:g1equadiff},\ref{eq2:r1equadiff}) satisfied
 by $R_0$ and $R_1$, one can show that $\widetilde{R}_0$ and $\widetilde{R}_1$
satisfy~:
\begin{eqnarray}
 \label{eq2:r0tildeqdiff}
\partial_t \wt{R}_0 &=&  \partial_x^2 \wt{R}_0  + \left( -v +2\frac{\partial_x
Q_e(x,t+t')}{Q_e(x,t+t')} \right) \partial_x \wt{R}_0  \\
& &+ \sum_{k=2}^\infty
\beta_k Q_e(x,t+t')^{k-1} \left( \wt{R}_0^k - \wt{R}_0\right) \nonumber \\
\label{eq2:r1tildeqdiff}
\partial_t \wt{R}_1 &=&  \partial_x^2 \wt{R}_1  + \left( -v +2\frac{\partial_x
P_1(x,t+t')}{P_1(x,t+t')} \right) \partial_x \wt{R}_1 \\
& &+ \sum_{k=2}^\infty k
\beta_k Q_e(x,t+t')^{k-1} \left( \wt{R}_0^{k-1} \wt{R}_1- \wt{R}_1\right)\nonumber
\end{eqnarray}
(where we have used the fact that $Q_e(x,t)=P_0(x,t)$). The initial conditions (\ref{eq2:p1r1initial})
give $\wb{R}_0(x,0,t' ; f)= \wb{R}_1(x,0,t';f)= e^{-f(x)}$.

We are now going to show that the functions $\wt{R}_0$ and $\wt{R}_1$ can be
interpreted as the generating functions at time $t$ of a modified process
defined on $[0,T]$ with $T=t+t'$. To do so, we consider the following process for $0<t<T$ in the frame of the wall~:
\begin{itshape}
\begin{itemize}
 \item the system consists of a single particle of type $A_1$ (the "spine" particle) and an
arbitrary number of particles $A_0$;
 \item a particle of type $A_0$ diffuses at time $t$ in the frame of the wall with a drift
\begin{equation}
 v_0(x,t,T)= -v+2\frac{\partial_x Q_e(x,T-t)}{Q_e(x,T-t)}
\end{equation} and branches into $k$
particles of type $A_0$ at rate $\beta_k^{(0)}(x,t,T)=\beta_k
Q_e(x,T-t)^{k-1}$~:
\begin{equation}
 A_0 \xrightarrow{\beta_k^{(0)}(x,t,T)} k A_0
\end{equation}
 \item the particle of type $A_1$ diffuses in the frame of the wall with a drift
\begin{equation}
\label{eq:driftv1}
v_1(x,t,T)=-v+2\frac{\partial_x P_1(x,T-t)}{P_1(x,T-t)}
\end{equation} and branches
into one
particle of
type $A_1$ and $k-1$ particles
of type $A_0$ at rate $\beta_k^{(1)}(x,t,T) = k \beta_k
Q_e(x,T-t)^{k-1}$~:
\begin{equation}
 A_1 \xrightarrow{\beta_k^{(1)}(x,t,T)} A_1+
(k-1) A_0
\end{equation}
\end{itemize}
\end{itshape}
One can notice that now the drifts and the mutation rates
depend both on space and time through the functions $Q_e(x,t)$ and $P_1(x,t)$. Moreover, as $P_1(0,t)=0$ and $\partial_xP_1(0,t)
\neq 0$, the particle $A_1$ is never absorbed by the wall (whenever
it approaches the wall, it is
pushed away (\ref{eq:driftv1}) from it by the drift $\partial_x P_1/P_1$ which diverges when
$x\to 0$).
The extinction probability $Q_e^{(0)}(x,t ; t')$ of a particle $A_0$ is given by $Q_e^{(0)}(x,t ; t') = Q_e(x,t)/Q_e(x,t+t')$ and one can verify that it satisfies (\ref{eq2:r0tildeqdiff})~: indeed it is equal to $\widetilde{R}_0(x,t,t';\infty)$. Thus, at $t'=0$, i.e. at the final time $T$, all particles $A_0$ have disappeared.

For this modified process, one can consider the new generating
function  $\wb{G}^{(0)}(x,t,T,f) = \langle \exp( -\sum_i f(x_i^{(t)})) \rangle$ (resp.
$\wb{G}^{(1)}(x,t,T,f)$) where the summation at time $t$ is over both particles $A_0$ and $A_1$, given that we start with a single
initial
individual of type $A_0$ at position $x$ (resp. $A_1$) at $t=0$. In this
modified process, the offspring are still independent as in
(\ref{eq2:Qe_pde_deriv})
and the function $\wb{G}^{(0)}$ (resp. $\wb{G}^{(1)}$) is the product of the
$\wb{G}^{(i)}$'s of the offspring of the initial individual of type $A_0$ (resp.
$A_1$). As in (\ref{eq2:Qe_pde_deriv}), one can split the time interval
$[0,T+dt]$ into two intervals
$[0,dt]$ and $[dt,T+dt]$ and one gets~:
\begin{equation}
 \begin{split}
  \wb{G}^{(0)}&(x,t+dt,T+dt,f) =\\
& \int \frac{e^{-\eta^2/4}}{\sqrt{4\pi}}
\wb{G}^{(0)}(x+\eta \sqrt{dt} + v_0(x,0,T+dt) dt,t,T,f) d\eta \\
 & +\sum_k \beta_k Q_e(x,T+dt)^{k-1} \left( \wb{G}^{(0)}(x,t,T,f)^k
-\wb{G}^{(0)}(x,t,T,f) \right)
 \end{split}
\end{equation}
and, as $dt$ is infinitesimal, $\wb{G}^{(0)}$ has the same evolution (\ref{eq2:r0tildeqdiff}) as $\wb{R}_0$. Similarly, writing the evolution of $\wb{G}^{(1)}$ in the same way shows that it has the same evolution as  $\wb{R}_1$. Since the $\wb{G}^{(0)}$ and $\wb{R}_0$ (resp. $\wb{G}^{(1)}$ and $\wb{R}_1$) have also the same initial conditions, one has $\wb{R}_0(x,t,t';f)=\wb{G}^{(0)}(x,t,T,f)$ and $\wb{R}_1(x,t,t';f)=\wb{G}^{(1)}(x,t,T,f)$ and the $\wb{R}_i$'s introduced in (\ref{eq2:linkg2conditionned-defRtilde}) can
be interpreted as the generating functions of a modified process defined on
$[0,T]$.


When
$T\rightarrow\infty$ and $v<v_c$, the extinction probability $Q_e(x,T-t)$
converges to $Q_e^*(x)\neq 0$ and $P_1(x,T-t)$ decreases as in (\ref{eq:largeRmPm}). Thus, as $T\to\infty$, the drifts and the branching rates of
the modified
process become independent of $t$~:
\begin{subequations}
\label{eq:modproc:Tinfty}
\begin{eqnarray}
  v_0(x,t,T)&\to& w_0(x)=  -v +2
\frac{\partial_x Q_e^*(x)}{Q_e^*(x)}, \quad \beta_k^{(0)}(x,t,T) \to
\beta_k Q_e^*(x)^{k-1} \\
 v_1(x,t,T)  &\to & w_1(x)= -v +  2
\frac{\partial_x \phi_1(x)}{\phi_1(x)}, \quad\beta_k^{(1)}(x,t,T)
\to
k\beta_k Q_e^*(x)^{k-1}
\end{eqnarray}
\end{subequations}

From the expressions (\ref{eq:modproc:Tinfty}), one can understand the best
strategy
for the system to have a single survivor at $T=t+t'$. The
branching rates decrease as $x$ increases and vanish as $Q_e^*(x)$
goes to $0$~: no population can develop in the region where $Q_e^*(x) \simeq 0$
and it prevents the population from growing exponentially (which would not be
compatible with a finite size at $T$). The particle $A_1$ cannot be absorbed because $v_1(x,t,T)\to \infty$ as $x\to 0$.

\bigskip

\textbf{Remark}~: for the birth-death process of section \ref{sec:toy-continuous}, one could similarly construct a modified process \cite{hardyharris} with branching rates $\beta_k {Q_e^*}^{k-1}$ for particles $A_0$ and $k\beta_k
{Q_e^*}^{k-1}$ for the \emph{spine} $A_1$, to describe the quasi-stationary regime.

\textbf{Remark}~: The construction of a modified process can be easily adapted
to intermediate
regimes $0<t<T$ conditioned on the survival of two (or more generally to $p$)
survivors at $T$. The differential equations satisfied by
$R_2$ and
$P_2$ coud in the same way be interpreted as the dynamics of a system of three types
of particles with the following branching rates~:
\begin{eqnarray*}
 A_0 \to k A_0, & & \text{rate $\beta_k Q_e(x,T-t)^{k-1}$}   \\
 A_1 \to A_1 + (k-1) A_0, & & \text{rate $k\beta_k Q_e(x,T-t)^{k-1}$}
\\
A_2 \to A_2 + (k-1) A_0, & &\text{rate $k\beta_k Q_e(x,T-t)^{k-1}$}
\\
A_2 \to A_1 + A_1, & &\text{rate $\frac{k(k-1)}{2}\beta_k
Q_e(x,T-t)^{k-2}P_1(x,T-t)$}
\end{eqnarray*}
where $A_0$ and $A_1$ have the same drifts as before and $A_2$ has an
additional drift $2\partial_x P_2(x,T-t)/P_2(x,T-t)\simeq
2\partial_x\phi_1(x)/\phi_1(x)$ for $T\to\infty$.
The exponential decay of $P_1$ to $0$ with a relaxation time $\tau_1$
shows that the reaction $A_2\to A_1+A_1$ occurs only for $t \simeq T$.

\subsection{Average profiles}
\label{subsec:modproc:profile}
The average density profile $\rho(X,t)$ of the population is defined such that $\rho(X,t)dX$ is the
average number of individuals located in the interval $[X,X+dX]$ in the moving
frame of the wall. It is easy to see that it satisfies~:
\begin{equation}
 \label{eq:profileequadiff-nocond}
\partial_t \rho = \partial_X^2 \rho + v \partial_X\rho + \textstyle{\sum_k} \beta_k
(k-1) \rho
\end{equation}
with $\rho(0,t)=0$ because of the absorption by the wall and
$\rho(X,0)=\delta(X-x)$ where $x$ is the position of the initial
individual. If one introduces the growth rate $\bc=
\sum_k\beta_k(k-1)$, the solution is given by~:
\begin{equation}
\label{eq:profile-nocond}
 \rho(X,t) = \frac{1}{\sqrt{4\pi t}} e^{ (\bc-v^2/4) t} e^{-v(X-x)}
\left[ e^{ -(X-x)^2/4t} - e^{ -(X+x)^2/4t} \right]
\end{equation}
Similar expressions were obtained in \cite{antal} for a model of cell proliferation.

If $v > v_c=2{\bc}^{1/2}$, the average density decreases to zero and this corresponds to
an almost sure extinction of the population. For $v<v_c$, the divergence corresponds to the
exponential growth whenever the population survives.

\bigskip

We are now going to calculate the average
profile $\rho_\text{qs}(X)$ in the quasi-stationary regime using the modified process of section \ref{subsec:modifiedprocess}. There are two contributions to this profile~:
\begin{equation}
\label{eq:link:rhoqs:rho1:rho0}
 \rhoqs(X) =  \rho_{1,\text{st}}(X) + \rho_{0,\text{st}}(X)
\end{equation}
where $\rho_{1,\text{st}}(X)$ and $\rho_{0,\text{st}}(X)$ are the stationary
average density profiles of the particles $A_1$ and $A_0$ in the modified
process. In the moving frame of the wall and in the stationary regime,
particles of type $A_1$ (resp.
$A_0$) have a drift
$w_1(X)=-v+2\partial_x \phi_1/\phi_1$ (resp. $w_0(X)=-v+2\partial_x
Q_e^*/Q_e^*$). The average density profiles $\rho_0$ and $\rho_1$ satisfy equations similar to (\ref{eq:profileequadiff-nocond}) where we use now the drifts and the branching rates (\ref{eq:modproc:Tinfty}) of the modified process~:
\begin{eqnarray}
 \partial_t \rho_0(X,t) &=& \partial_X^2 \rho_0(X,t)  - \partial_X\left(
w_0(X)\rho_0 (X,t)\right) \nonumber \\
& & + \sum_k (k-1) \beta_k Q_e^*(X)^{k-1}  \rho_0(X,t) \nonumber\\
& &+\sum_k k(k-1) \beta_k Q_e^*(X)^{k-1}  \rho_1(X,t)  \label{eq:evolrho0}\\
\partial_t \rho_1(X,t) &=& \partial_X^2 \rho_1(X,t)  - \partial_X\left(
w_1(X)\rho_1 (X,t)\right) \label{eq:evolrho1}
\end{eqnarray}
There is no source term for $\rho_1$ since the number of particles $A_1$ is
conserved. As there is initially a single particle $A_1$, the density $\rho_1$ is the distribution of its position at time $t$. The
stationary profile $\rho_{1,\text{st}}$ is directly obtained from the
expression of $w_1(X)$ and (\ref{eq:evolrho1})~:
\begin{equation}
\label{eq:rho1statio}
 \rho_{1,\text{st}}(X) =  C \phi_1(X)^2 e^{-vX}
\end{equation}
where $C$ is a normalization constant such that $\int_0^\infty
\rho_{1,\text{st}}(X) dX= 1$. The density $\rho_{1,\text{st}}$ depends only on
the slowest eigenvector $\phi_1$ and its shape near
the critical velocity will be studied in section \ref{sec:eigenvalues}.

The extinction probability of $A_0$ particles in the modified process is given by construction by $\tilde{R}_0(x,t,
t';\infty)=Q_e(x,t)/Q_e(x,t+t')$ and thus becomes $Q_e(x,t)/Q_e^*(x)$ as $t'\to\infty$ in the quasi-stationary regime. Therefore a particle $A_0$ gets extinct with probability $1$ in the long time limit. Thus particles $A_0$ are produced by the particle $A_1$ and later on  are absorbed by the wall. Their stationary density $\rho_{0,\text{qs}}$ satisfies~:
\begin{align}
\label{eq:qsprofile:rho0:ode}
 \partial_X^2
\rho_{0,\text{st}}(X)  - &\partial_X\left(
w_0(X)\rho_{0,\text{st}} (X)\right) +\sum_k \beta_k Q_e^*(X)^{k-1} (k-1)
\rho_{0,\text{st}}(X) \\
& = - g''\left( Q_e^*(X)\right) Q_e^*(X)
\rho_{1,\text{st}}(X) .\nonumber
\end{align}
The change of variable $\rho_{0,\text{st}}(X)= e^{-vX} Q_e^*(X) \psi(X)$
leads for $\psi$ to an equation of the type~:
\begin{eqnarray}
\label{eq:qsprofile:psi:genericode}
 \Lin[\psi] &=& K(X) \\
 K(X) &=& - g''\left( Q_e^*(X)\right)
\rho_{1,\text{st}}(X) e^{vX} \label{eq:def:K}
\end{eqnarray}
where $\Lin$ is the linear operator defined in (\ref{eq2:Qe:linop:def}).
This equation is an inhomogeneous second order linear differential equation that can be solved
easily (because one solution of the homogeneous equation, $\partial_x Q_e^*(x)$, is known). The general solution of
(\ref{eq:qsprofile:psi:genericode}) is
\begin{equation*}
\label{eq:qsprofile:psi:genericsol}
\begin{split}
 \rho_{0,\text{st}}(X) = &e^{-vx} Q_e^*(X) \partial_x Q_e^*(X) \\
&\times \int_{a}^X
\frac{dy}{[\partial_x
Q_e^*(y)]^2 e^{-vy}} \int_{b}^{y} \partial_x Q_e^*(z) e^{-vz} K(z) dz.
\end{split}
\end{equation*}
The stationary profile $\rho_{0,\text{st}}(X)$ is
then obtained by choosing $a=0$ and
$b\to\infty$ for $\rho_{0,\text{st}}(X)$ to vanish at $X=0$ and $X\to\infty$~:
\begin{equation}
 \label{eq:qsprofile:rho0:exactresult}
\begin{split}
\rho_{0,\text{st}}(X) = &e^{-vX} Q_e^*(X)\partial_x Q_e^*(X) \\
& \times \int_{0}^X
\frac{dy}{[\partial_x Q_e^*(y)]^2 e^{-vy}}  \int_y^\infty \partial_x
Q_e^*(z)  g''\left( Q_e^*(z)\right) \rho_{1,\text{st}}(z) dz.
\end{split}
\end{equation}

Equations (\ref{eq:link:rhoqs:rho1:rho0}, \ref{eq:rho1statio},
\ref{eq:qsprofile:rho0:exactresult}) give the exact average quasi-stationary profile
in terms of only $Q_e^*$ and $\phi_1$ which have been used for
the construction of the quasi-stationary state. Section \ref{sec:universality}
will describe the limit $v\to v_c$ of $Q_e^*(x)$, $\phi_1(x)$ and $\rhoqs(X)$.

The results (\ref{eq:link:rhoqs:rho1:rho0}, \ref{eq:rho1statio},
\ref{eq:qsprofile:rho0:exactresult}) could also be derived from the dynamical
system approach of section \ref{subsec:dynamical}. The average
quasi-stationary profile $\rho_\text{qs}(X)$ at distance $X$ from the wall is
obtained by
considering the test function $f(x)=\mu\delta(x-X)$ and taking the
$\mu$-derivative of (\ref{eq:linkA1qs}) at $\mu=0$, as for the average size in
(\ref{eq:Nmoy-A}). Relations (\ref{eq2:AF1}, \ref{eq2:AF2}) between
$\A^{(1)}[\phi_i]$, $\A^{(2)}[\phi_i,\phi_j]$, $\Lin$ and
$\F^{(2)}[\phi_i,\phi_j]$ are then  exactly equivalent to the differential
equations (\ref{eq:evolrho0},\ref{eq:evolrho1}) in their stationary regime.


\section{Universality near in the critical velocity}
\label{sec:universality}

In this section, we determine the scaling properties of the quasi-stationary profile when $v_c -v \to 0^+$. We are going to see that, in this limit, everything can be expressed in terms of the shape $Q_{v_c}(x)$ of a front on the infinite line at the critical velocity.

\subsection{Extinction probability and relaxation times}
\label{sec:eigenvalues}

The stable fixed point $Q_e^*(x)$ of the functional $\F$ defined in
(\ref{eq2:deffunctionalF}) satisfies the differential equation
(\ref{eq:defQestar}) 
with the boundary condition $Q_e^*(0)=1$ and $Q_e^*(x)$ goes to $0$ or $1$ as
$x\to \infty$.

In appendix \ref{appendix:eigenvalues}, we use a perturbation theory to analyze (\ref{eq:defQestar}) on the infinite line for $v_c-v \to 0^+$ to obtain the extinction probability $Q_e^*(x)$ in presence of the wall in the same limit. The solution $Q_{v_c}$ of (\ref{eq:defQestar}) on the infinite line such that $Q_{v_c}(x_0)=1/2$ (where $x_0$ is some arbitrary position), $Q_{v_c}(x)\to 1$ as $x\to -\infty$ and $Q_{v_c}(x) \to 0$ as $x\to +\infty$ is given for large negative $x$~:
\begin{equation}
\label{eq:qvcshape}
 Q_{v_c} (x)= 1 + (A_c x +B_c) e^{v_c x/2} + O\left( e^{v_c x} \right)
\end{equation}
where $A_c$ and $B_c$ are constant depending on $x_0$.
For $v$ close $v_c$, one can show (see appendix \ref{appendix:eigenvalues}) that, on a domain of length $L$ that we will call region I, $Q_e^*(x)$ is given by~:
\begin{equation}
\label{eq:Qestar:lindomain}
 Q_e^*(x) \underset{\text{region I}}{\simeq} 1 - \frac{A_c L}{\pi}   \sin\left( \frac{\pi x}{L} \right) \exp\left[ \frac{v_c}{2}(x-L-B_c/A_c)\right]
\end{equation}
where $A_c$ and $B_c$ are the constants defined as in (\ref{eq:qvcshape}). The length $L$ is given by~:
\begin{equation}
 \label{eq:def:length:L}
 L=2\pi/\sqrt{v_c^2-v^2} \underset{v\to v_c}{\sim} (v_c-v)^{-1/2}
\end{equation}
and diverges at the critical velocity. In the domain $x \gtrsim L$ (region II), the non-linearities of (\ref{eq:defQestar}) must be taken into account and $Q_e^*$ is given at leading order by~:
\begin{equation}
 Q_e^*(x) \underset{\text{region II}}{=} Q_{v_c}(x-L- B_c/A_c) + O(1/L^2)
\end{equation}

The relaxation times $\tau_n$ of the dynamics (\ref{eq2:Qe_pde}) near the
stable fixed point $Q_e^*$ are related to the eigenvalues $\lambda_n=-1/\tau_n$ of
the linear operator (\ref{eq2:Qe:linop:def}). An eigenvector $\phi_n$ with eigenvalue $\lambda_n$ satisfies the second
order differential equation~:
\begin{equation}
 \partial_x^2 \phi_n -v \partial_x \phi_n + g'\left( Q_v(x) \right)
\phi_n = \lambda_n \phi_n
\end{equation}
with the boundary conditions~:
\begin{equation}
 \label{eq:eigenboundary}
 \phi_n(0) = 0, \quad \phi_n(x) \xrightarrow{x\to +\infty} 0
\end{equation}

As for $Q_e^*$, the shape of the eigenvectors $\phi_n$ can be obtained from the ones on the infinite line as shown in appendix (\ref{appendix:eigenvalues}). The effect of the boundary condition (\ref{eq:eigenboundary}) on the wall is to select the ones which vanish at $x=0$. A perturbative expansion in $v_c-v$ and $\lambda$ shows that the first  eigenvalues $\lambda_n$ in presence of the absorbing wall are given by~:
\begin{equation}
\label{eq:lambdan}
 \lambda_n \simeq - \frac{ (n^2-1)\pi^2 }{L^2} - \frac{ 4 n^2 \pi^2   }{ v_c L^3}
- \frac{ 12 n^2 \pi^2 }{ v_c^2 L^4}  + O \left( \frac{1}{L^5} \right)
\end{equation}
One can notice that the first three terms in this expansion are independent of the precise form $g(Q)$ of the non-linearities.
The eigenvector $\phi_n$ associated to $\lambda_n$ is given (up to a multiplicative constant) at leading order by~:
\begin{subequations}
 \label{eq:phishape}
\begin{equation}
\label{eq:phishapelin}
 \phi_n(x) \underset{\text{region I}}{\simeq}  \frac{A_c v_c L (-1)^{n-1}}{2 n\pi} \sin\left( \frac{n \pi x}{L} \right) \exp\left[ \frac{v_c}{2} \left(x-L- \frac{B_c}{A_c}\right)\right]
\end{equation}
in the region I of size $L$ near the wall and by~:
\begin{equation}\label{eq:phishapenonlin}
 \phi_n(x) \underset{\text{region II}}{\simeq} \partial_x Q_{v_c}(x-L-B_c/A_c)
\end{equation}
\end{subequations}
in the region II such that $x \gtrsim L$. Appendix \ref{appendix:eigenvalues} also gives the order of magnitude of the first correction to $\phi_n$ in both regimes $x\ll L$ and $x>L$. As the relaxation times are given by $\tau_n=-1/\lambda_n$, the longest relaxation time $\tau_1$ is given (\ref{eq:lambdan}) by
\begin{equation}
 \label{eq:tau1}
\tau_1 \simeq  \frac{L^3 v_c}{4 \pi^2} \sim \frac{\pi}{\sqrt{2v_c}} (v_c-v)^{-3/2}
\end{equation}
and is much larger than all the other $\tau_n$'s ($n\ge 2$). One can notice that the same expression of $\tau_1$ was obtained in \cite{epl-derridasimon} by a very different approach.

\subsection{Quasi-stationary profile}

The quasi-stationary profile (\ref{eq:link:rhoqs:rho1:rho0}) is the sum of two
contributions (\ref{eq:rho1statio},\ref{eq:qsprofile:rho0:exactresult}) of
the modified process. For $v$ close to $v_c$, the normalization constant $C$ in (\ref{eq:rho1statio}) is dominated by the contribution of region I and is given at
lowest order by $C\simeq8\pi^2 e^{v_c(L+B_c/A_c)}/(A_c^2 v_c^2 L^3)$.
Thus
the stationary profile $\rho_{1,\text{st}}(X)$ at distance $X$ from the wall in
the modified process is given by~:
\begin{equation}
\label{eq:rho1stnearvc}
 \rho_{1,\text{st}}(X) \underset{\text{region I}}{\simeq} \frac{2}{L} \sin\left(\frac{\pi X}{L}\right)^2
\end{equation}
for $X<L$. For $X\gtrsim L$ (region II), the density is given by~:
\begin{equation}
\rho_{1,\text{st}}(X) \underset{\text{region II}}{\simeq} \frac{ 8\pi^2 e^{v_c(L+B_c/A_c)}}{A_c^2 v_c^2 L^3}e^{-vX} \left[\partial_xQ_{v_c}(X-L- B_c/A_c)\right]^2
\end{equation}
The corrections to this profile are of order $1/L$ in both regions.

The density profile $\rho_{0,\text{st}}(X)$ given by
(\ref{eq:qsprofile:rho0:exactresult}) can be rewritten in the following way~:
\begin{equation}
\label{eq:rho0st:defI}
 \rho_{0,\text{st}}(X) = e^{-vX} Q_e^*(X)\partial_x Q_e^*(X) \int_{0}^X
\frac{dy}{[\partial_x Q_e^*(y)]^2 e^{-vy}} \left( I_\infty - I(y) \right)
\end{equation}
where
\begin{equation}
 I(y) = \int_0^y \partial_x Q_e^*(z) g''( Q_e^*(z))
\rho_{1,\text{st}}(z) dz \label{eq:rho0st:Iyint}
\end{equation}
and $I_\infty= \lim_{y\to+\infty} I(y)$.
For $x<L$, $\phi_1(x)$ and $\partial_x Q_e^*$ are of
order $e^{v_c (x-L)/2}$ and $Q$ is of order $1$~; for $x-L \gg 1$ (region II), all the
terms in the integral (\ref{eq:rho0st:Iyint}) decrease exponentially. Thus, the integral
is dominated by the region of size of
order 1 near $x\simeq L$ and $I_\infty$ scales for $v_c-v\to 0^{+}$ as~:
\begin{equation}
\label{eq:Iinftyscaling}
 I_\infty \simeq \frac{1}{L^3} \left[ \frac{8\pi^2 }{A_c^2 v_c^2} \int_{-\infty}^{+\infty} [\partial_x Q_{v_c}(z)]^3 g''( Q_{v_c}(z)) e^{-v_c z} dz\right]
\end{equation}
One notices that $I_\infty$ is negative since $Q_e^*(x)$ is a decreasing function and that $I_\infty$ is independent of the choice of the reference $x_0$ such that $Q_{v_c}(x_0)=1/2$.
On the other hand, for the same reasons, $I(y)$ is exponentially small and can be neglected compared to $I_\infty$ in the linear region $0 < y<L$.

At leading order in the region I, one gets from (\ref{eq:rho0st:defI})~:
\begin{equation}
\label{eq:rho0scaling1}
 \rho_{0,\text{st}}(X) \simeq I_\infty e^{-v_c X} \partial_x Q_e^*(X) \int_0^X \frac{1}{[\partial_x Q_e^*(y)]^2 e^{-vy}} dy
\end{equation}
From (\ref{eq:Qestar:lindomain}), $\partial_x Q_e^*(x)$ is given for $x<L$ by~:
\begin{equation*}
 \partial_x Q_e^*(x) \underset{\text{region I}}{\simeq} -\frac{A_c L}{\pi} \left[ \frac{v_c}{2} \sin\left( \frac{\pi x}{L}\right) + \frac{\pi}{L} \cos\left( \frac{\pi x}{L} \right) \right] e^{\frac{v_c}{2} (x-L-B_c/A_c)}
\end{equation*}
up to exponentially small corrections. The integral in (\ref{eq:rho0scaling1}) is equal to~:
\begin{eqnarray*}
\int_0^X \frac{1}{[\partial_x Q_e^*(y)]^2 e^{-vy}} dy &\simeq&
\frac{\pi^2 e^{v_c(L+B_c/A_c)}}{A_c^2 L^2} \int_0^X \frac{dy}{ \left[ \frac{v_c}{2} \sin\left( \frac{\pi y}{L}\right) + \frac{\pi}{L} \cos\left( \frac{\pi y}{L} \right) \right]^2} \\
&\simeq& \frac{e^{v_c(L+B_c/A_c)}}{A_c^2} \frac{ \sin\left( \frac{\pi X}{L} \right)}{\frac{v_c}{2} \sin\left( \frac{\pi X}{L}\right) + \frac{\pi}{L} \cos\left( \frac{\pi X}{L} \right)}
\end{eqnarray*}
Finally the density of $A_0$ particles is given in the region I by~:
\begin{equation}
\rho_{0,\text{st}}(X)\simeq  -\frac{I_\infty L e^{\frac{v_cB_c}{2A_c}}}{A_c \pi} \sin\left(\frac{\pi X}{L}\right) \exp\left[ -\frac{v_c}{2} (X-L)\right]
\label{eq:rho0stnearvc}
\end{equation}
In the region $x<L$, the number of particles $A_0$ is exponentially large and $\rho_{1,\text{st}}(X)$ is bounded and much smaller than $ \rho_{0,\text{st}}(X) $  as shown in
(\ref{eq:rho1stnearvc}, \ref{eq:rho0stnearvc}). Thus the quasi-stationary
average profile $\rho_{\text{qs}}(X)$ scales in the linear domain as~:
\begin{equation}
\label{eq:qsprofileaverage}
 \rho_{\text{qs}}(X) \underset{\text{region I}}{\simeq}  \frac{ K v_c^2}{4 \pi L^2 } e^{-v_c
(X-L)/2} \sin\left(\frac{\pi X}{L}\right)
\end{equation}
with a constant $K$ given by~:
\begin{equation}
\label{eq:defconstantK}
K = -\frac{32 \pi^2 e^{ \frac{1}{2} v_c B_c /A_c}}{A_c^3 v_c^4} \int_{-\infty}^{+\infty}  [\partial_x Q_{v_c}(z)]^3 g''( Q_{v_c}(z)) e^{-v_c z} dz
\end{equation}
One can check that this expression of $K$ is independent of the reference point $x_0$ chosen for $Q_{v_c}$.
For $X-L$ of order $1$, $\rho_{\text{qs}}(X)$ is of order $1/L^3$ and in the domain $X>L$ (region II), the
density decreases exponentially, so that the total number of individuals is dominated by
the region $0<X<L$. The average size in the quasi-stationary regime is thus given by~:
\begin{eqnarray}
\label{eq:qssizeaverage}
\moyqs{N}&=& \int_0^\infty \rhoqs(X)dX \simeq  \frac{K}{L^3} e^{v_c L/2} \\
&\sim& (v_c-v)^{3/2} \exp\left[ \pi\sqrt{\frac{v_c}{2}} (v_c-v)^{-1/2} \right] \nonumber
\end{eqnarray}

The divergence (\ref{eq:qssizeaverage}) of
$\moyqs{N}$ can also be obtained from (\ref{eq:Nmoy-A}) by expanding $Q_e^*$ in
the basis of the eigenvectors and using (\ref{eq2:AF1}, \ref{eq2:AF2})~: the
leading term (\ref{eq:qssizeaverage}) could then be obtained by truncating the
expansion of $Q_e^*$ after the first eigenvectors $\phi_1$, as the first
eigenvalue $\lambda_1\simeq (v_c-v)^{-3/2}$ is much smaller than the next ones
that scale as $(v_c-v)^{-1}$.

In \cite{brunetgenea,brunetderridacoal}, Brunet et al. considered a
population whose size $N$ is kept constant by
always selecting the $N$ individuals with the largest $x_i$.
From \cite{brunetpheno}, they predicted that the effect of the finite size $N$
is to shift the average velocity $\moy{v}$ of the population by an amount~:
\begin{equation}
\label{eq:brunet:shiftv}
 \moy{v} = v_c - \frac{v_c \pi^2}{2 \log^2 N} + ( v_c \pi^2) \frac{3 \log\log
N}{\log^3 N} + \ldots
\end{equation}
\modif{This  $N$ dependence of the velocity was recently derived rigorously in \cite{muellermytnik}}.
It is interesting to notice that this relation between $\moy{v}-v_c$ and $N$ is the same as (\ref{eq:qssizeaverage}) between $\moy{N}$ and $v-v_c$ in the
quasi-stationary state of the present model with an absorbing wall.

\section{Exponential model}
\label{sec:exponential}

There is a simplified version of the problem of a branching random walk with an
absorbing wall, the exponential model, which can be solved exactly
\cite{brunetgenea,brunetderridacoal}. In this exponential model, time is discrete and at each generation,
each individual is replaced by its offspring
distributed according to a Poisson point process~: if the parent is at position
$x$, then for every interval $[y,y+dy]$ there is an offspring in this interval
with probability $\psi(y-x)dy$.

By analyzing the first time step one can show that the
extinction probability $Q_e(x,t)$ at time $t$ of the descendance of an individual located at distance $x$ from the wall at time $t=0$ evolves according to~:
\begin{equation}
\label{eqexp:recursionQ}
 Q_e(x,t+1)  =  \F( Q_e(x,t) )= \exp\lp   -\int_0^\infty \psi(y+v-x) \left(1
-Q_e(y,t)\right) dy \rp
\end{equation}
For a general $\psi$, $\F$ is a functional of $Q_e$ for which one could try to generalize the approaches of sections \ref{sec:conditionned} and \ref{sec:modified}. For the special case $\psi(x)=e^{-x}$, however, the problem reduces to the study of a
one-dimensional mapping. For any $Q_e(x,0)$, it is easy to see from
(\ref{eqexp:recursionQ}) that at $t\geq 1$ $Q_e(x,t)$ takes the form
\begin{equation}
\label{eqexp:Qe:reduction}
 Q_e(x,t)=e^{-c_t e^x}
\end{equation}
and depends on a single parameter $c_t$
which satisfies the following recursion~:
\begin{eqnarray}
\label{eqexp:recursionc}
 c_{t+1} &=&  F( c_{t}) \\
 F(c) &=& e^{-v} \left( 1- \int_0^\infty e^{-y} e^{-c e^y}
dy\right)\label{eqexp:recursionc-defF}
\end{eqnarray}
For all $v$, there is an attractive fixed point $c_*>0$ and the population
always has a non-zero survival probability 
$1-Q_e^*(x)=1-e^{-c_*e^{x}}$ in the long time limit. The relaxation of
$Q_e(x,t)$ to $Q_e^*(x)$ is entirely controlled by the relaxation of $c_t$ to
$c_*$~: there is a unique eigenvalue $\Lambda=F'(c_*)<1$ with the eigenvector
\begin{equation}
\phi_1(x)= -\exp( x-c_* e^{x})
\end{equation} and one has $
Q_e(x,t)\simeq Q_e^* (x) - K \Lambda^t e^{x-c_*e^{x}}$.
Generating functions $G_1$ and $G_2$ defined as in (\ref{eq:def:genefuncG1},\ref{eq:defgeneG2}) also satisfy the same dynamics (\ref{eqexp:recursionQ}). A general initial condition $Q(x,0)=Q_0(x)$ is mapped at the first type step to a function $Q(x,1)=e^{-c(Q_0)e^x}$ where the coefficient $c(Q_0)$ is defined by
\begin{equation}
c(Q_0) = e^{-v}\left( 1- \int_0^\infty e^{-y} Q_0(y)dy\right)
\end{equation}
and has the following long time behaviour~:
\begin{equation}
 Q(x,t)= \exp( -c_t e^x ) \simeq Q_e^* (x) - \At( c(Q_0) )  \Lambda^{t-1} e^{x-c_*e^{x}}
\end{equation}
with an amplitude $\At(c)$ similar to the function $\A$ of section \ref{sec:toy}. As in (\ref{eq1:relationA-F}), the function $\At(c)$ satisfies~:
\begin{equation}\label{eq:Atc-F}
 \Lambda \At (c)   = \At( F(c) ), \quad \At'(c_*)=1\end{equation}
with $F$ given in (\ref{eqexp:recursionc-defF}).

The reduction to a one-parameter family $\exp(-c e^{x})$ makes the analysis of this spatial model
similar to what we did in section
\ref{sec:toy}.

For example, the generating function of the number $N(X)$ of individuals at a distance $x$ greater than a given $X$ from the wall is given in the quasi-stationary regime by (\ref{eq2:quasistat-A}) with $\A(Q)= \At(c(Q))/\Lambda$ and $f(x)=\mu\theta_X(x)\equiv\mu\theta(x-X)$. For the exponential model, it can be reduced to a simple function of $\mu$~:
\begin{equation*}
 \moyqs{ e^{-\mu N(X)} } = \frac{1}{\Lambda} \frac{d}{ds} \At\left( c\left( (Q_e^*+s\phi) e^{-\mu \theta_X} \right) \right) \Big|_{s=0}
\end{equation*}
By decomposing $c\left( (Q_e^*+s\phi_1) e^{-\mu \theta_X} \right)$ in the following way~:
\begin{eqnarray}
c\left( (Q_e^*+s\phi_1) e^{-\mu \theta_X} \right) &=& e^{-v}\left( 1 - \int_0^\infty e^{-y} \left(Q_e^*(y)+s\phi_1(y)\right)dy \right) \nonumber  \\
& &+ e^{-v}(1-e^{-\mu}) \int_X^\infty e^{-y} \left(Q_e^*(y)+s\phi_1(y)\right)dy  \nonumber
\\
& =&  c_* + s\Lambda + (1-e^{-\mu}) \left(J_0(X)+ s\Lambda J_1(X)\right) \nonumber\\
 J_0(X) &=& e^{-v} \int_X^\infty e^{-y -c_*e^{y}} dy \label{eq:J0def} \\
 J_1(X) &=& \frac{1}{\Lambda}\int_X^\infty e^{-c_*e^y} dy \label{eq:J1def}
\end{eqnarray}
where one can verify that $J_1(0)=1$, one obtains the generating function~:
\begin{equation}
 \moyqs{ e^{-\mu N(X)} }=\big( 1 - (1-e^{-\mu}) J_1(X) \big) \At'\left( c_* + (1-e^{-\mu}) J_0(X) \right)
\end{equation}
It follows that  the average density profile $\moyqs{\rho(X)}$ obtained by taking the derivative of $\moyqs{N(X)}$ and the average size of the population are given by~:
\begin{eqnarray}
 \moyqs{\rho(X)} &=& -J_1'(X)  + \At^{(2)}(c_*)  J_0'(X) \label{eqexp:rhoqs}\\
\moyqs{N} &=& 1+ \left( -\At^{(2)}(c_*) \right) J_0(0) \label{eqexp:Nqs}
\end{eqnarray}

Let us now analyze the large $v$ limit (which for the exponential model plays the role \cite{brunetderridacoal} of the $v\to v_c$ limit of section \ref{sec:universality}). The mapping $F(c)$ has the following expansion near $c=0$ (corresponding to
$Q_e^*(x)=1$)~:
\begin{equation}
\label{eqexp:Fexpansion}
 F(c) = e^{-v} \lp  -c \ln c +(1-\gamma) c + \frac{c^2}{2} - \frac{c^3}{12}
+\ldots \rp 
\end{equation}
where $\gamma=-\Gamma'(1)$ is Euler's constant. For large $v$, the stable fixed
point $c_*$ of $F$ is given by 
$$c_{*}\simeq e^{1-\gamma} \exp(-e^{v})$$ and $\Lambda\simeq 1-e^{-v}$.
As in section \ref{sec:toy}, one can compute $\At^{(2)} \simeq -1/c_*$ from (\ref{eq:Atc-F}). The functions $J_0$ and $J_1$ in (\ref{eq:J0def},\ref{eq:J1def}) have different shapes for $X<L$ (region I) and $X>L$ (region II) where the length $L$ is given by~:
\begin{equation}
L=-\ln c_*\simeq e^{v}
\end{equation}
Thus the leading terms in (\ref{eqexp:rhoqs},\ref{eqexp:Nqs}) when $X<L$ (region I) correspond to $J_0$ contributions and, for $v\to \infty$, the average quasi-stationary profile and sizes are given by~:
\begin{eqnarray}
 \moyqs{ \rho(X)} &\simeq& \frac{1}{L} e^{-(X-L)} \quad\text{(region I, $X<L$)}\\
 \moyqs{ N} &\simeq&  \frac{e^{-v}}{c_*} \sim \frac{1}{L} e^{L} \label{eqexp:Nmoyvinfty}
\end{eqnarray}
In region II (for $X>L$), $\moyqs{ \rho(X)}$ is decreasing as $e^{-c_*e^X}$.
As in section \ref{sec:universality}, the region of length $L$ near the wall has the dominant contributions to the size of the population. As in section \ref{sec:modified}, one could interpret the $J_1$ terms as the contribution of the $A_1$ particle and the $J_0$ terms as contributions of the $A_0$ particles in the context of the modified process.

One can notice that, in the exponential model too, (\ref{eqexp:Nmoyvinfty}) yields the same relation between the size of the population and the velocity as (39) in \cite{brunetderridacoal} (where the size of the population is kept constant but the velocity fluctuates).

In the limit $v\to \infty$, it is possible to obtain the whole generating function $\moyqs{e^{-\mu N}}$. In this limit, $N$ scales as $1/c_*$ and one should take $\mu$ of order $c_*$ and thus $\At(c)$ needs to be known for $c-c_* = O(c_*)$. If $F(c)$  in (\ref{eqexp:Fexpansion}) is truncated after the first two leading terms  near $c_*$ as in section \ref{subsec:toy:univ} and appendix \ref{appendix:heavytail}, one obtains $\At(c_* + c_* u) \simeq \ln(1+u)$ in the limit $v\to \infty$ and the following generating function of the size~:
\begin{equation}
\label{eqexp:sizeexpo}
 \Moyqs{ e^{-\mu N/\moyqs{N}}} \underset{v\to \infty}{\simeq} \frac{1}{1+\mu}
\end{equation}
Thus in the quasi-stationary regime the size of the population has an exponential distribution in contrast to what happens (\ref{eq1:N:univ}) in the Galton-Watson process. In particular, the ratio $\moyqs{N^2}/\moyqs{N}^2$ goes to $2$~: this result is similar to numerical results for lattice branching random walks obtained in \cite{epl-derridasimon}.

\section{Conclusion}

In the present work, we have studied the quasistationary regime of a branching
random walk in presence of an absorbing wall below the critical velocity 
\cite{epl-derridasimon}. To do so, we have developed two methods. The first one, discussed in section \ref{sec:conditionned}, is a dynamical system approach allowing to determine the generating functions of the size of the population.

In the second approach, one constructs from the
original process a modified stochastic process equivalent to the
quasi-stationary regime (section \ref{sec:modified}). This construction requires the knowledge of
$P_0(x,t)=Q_e(x,t)$ and $P_1(x,t)$, the probabilities of observing 
$0$ and $1$ survivor at time $t$ for an initial individual at $x$. These methods allowed us to
determine (\ref{eq:qsprofileaverage},\ref{eq:defconstantK},\ref{eq:qssizeaverage}) the average population size and the density profile in the quasi-stationary regime
near $v_c$. The average profile has a universal shape (\ref{eq:qsprofileaverage}) in a domain of size $L\sim
(v_c-v)^{-1/2}$ and the average size diverges as in (\ref{eq:qssizeaverage}) in a universal way which does not depend on the precise form of the branching rates $\beta_k$. These quantities have also been obtained for the exponential model in
section \ref{sec:exponential}. We noticed that both for the branching random walk and the exponential model, the relation between the velocity and the size of the population (\ref{eq:qssizeaverage},\ref{eqexp:Nmoyvinfty}) is the same as in a model recently studied in \cite{brunetderridacoal} for a population of constant size.

Beyond the average quasistationary profile which has a universal shape near $v_c$ (i.e. is independent of the precise form of the non-linearities $g(q)$ in (\ref{eq2:defnonling})), it
would be interesting to see whether the whole quasi-stationary measure is universal when $v\to v_c$. In particular, it would be interesting to know whether the size of the population has an exponential distribution in the quasi-stationary regime as in the exponential model (\ref{eqexp:sizeexpo}).

Our approach was limited to the case $v<v_c$. For $v>v_c$ one expects \cite{epl-derridasimon} that there is no quasi-stationary regime. The construction of the modified process of section \ref{sec:modified} remains however valid and this could be a starting point to understand the absence of a quasi-stationary regime.

More generally, the construction of the modified process is valid as long as
the individuals are independent. In particular, a modified process could be constructed to
describe the evolution of a population conditioned to a finite size $N_T=1$ in
non-uniform media, as in \cite{berestyckispeed,berestycki1}, where the branching
rates, the drift and the diffusion coefficients depend on the position $x$ and
where the domains have more complicated geometries ($d$-dimensional with
absorption on the boundaries). This could be a way of understanding the effect of
the geometry of the domain on the existence and on the properties of a quasi-stationary regime.

Using the modified process, one could also study the dynamical properties and the correlations in time of the quasi-stationary regime. In particular, one could try to determine the statistical properties of the genealogies in this quasi-stationary regime and compare them with the results obtained recently for a population of fixed size \cite{brunetgenea,brunetderridacoal}.

\appendix

\section{Universal distributions of population sizes in the
quasistationary regime near the transition in birth-death processes~: the case
of slowly decreasing branching rates}
\label{appendix:heavytail}

In this appendix, we discuss briefly the quasi-stationary regime of the
Galton-Watson process when the branching rates $\beta_k$ decay too slowly for $\overline{k^2}$
to be finite. We consider the case where for large $k$
\begin{equation}
 \beta_k \simeq B k^{-1-\eta}
\end{equation}
with $1 <\eta <2$. Then, for $Q$ close to $1$, $F(Q)$ takes the form for $Q$
close to $1$~:
\begin{equation}
F(Q) \simeq \alpha (Q-1) + C (1-Q)^\eta + \ldots
\end{equation}
with $C=B \Gamma(-\eta)$ and $\alpha=\sum_k (k-1)\beta_k$.

For $\alpha < 0$, the fixed point $Q_e^*=1$ is attractive and one gets from
(\ref{eq1:A-intrep}) for $Q$ close to $1$~:
\begin{equation}
 \A(Q) \simeq \frac{ (Q-1)}{ \left( 1- \frac{C}{\alpha}
(1-Q)^{\eta-1}\right)^{1/(\eta-1)}}.
\end{equation}
For small $\alpha>0$, the fixed point becomes $1-Q_e^*\simeq
(\alpha/C)^{1/(\eta-1)}$, the eigenvalue $F'(Q_e^*)=\lambda$ scales as
$\lambda\simeq -\alpha( \eta-1)$ and by integrating (\ref{eq1:A-intrep}), one
gets~:
\begin{equation}
 \A(Q) \simeq (1-Q)^{1-\eta} \left( 1-\frac{C}{\alpha} (1-Q)^{\eta-1}\right)
\frac{1}{\eta-1} \left(\frac{\alpha}{C}\right)^{\frac{\eta}{\eta-1}} .
\end{equation}
Then by taking $\mu$ small in (\ref{eq1:quasistat-A}), one gets for $\alpha
<0$~:
\begin{equation}
\label{eq:heavy:univ:neg}
 \moyqs{ e^{-\mu N}  } \simeq \left( \frac{1}{1-\frac{C}{\alpha}
\mu^{\eta-1}}\right)^{\frac{\eta}{\eta-1}}
\end{equation}
whereas for $\alpha > 0$, one gets~:
\begin{equation}
\label{eq:heavy:univ:pos}
 \moyqs{e^{-\mu N} } \simeq \left( 1+\mu
\left(\frac{C}{\alpha}\right)^{1/(\eta-1)} \right)^{-\eta}
\end{equation}

Expressions (\ref{eq:heavy:univ:neg},\ref{eq:heavy:univ:pos}), valid for $|\alpha|$ small, show that both
below and above the transition, the distributions of $N$ in the quasi-stationary
regime are universal (as up to a rescaling they depend only on $\eta$) and that they differ from what was found (\ref{eq1:N:univ})
in the case $\eta>2$).

\section{Perturbative calculation of the shape of the front, of the eigenvalues and eigenvectors for $v$ close to $v_c$}

In this appendix we determine perturbatively, for $v-v_c$ small,  the stable solution $Q_e^*$ of
(\ref{eq:defQestar}), the eigenvalues $\lambda_n$ and their associated eigenvectors  $\phi_n$ of the linear operator  $\Lin$ defined in (\ref{eq2:Qe:linop:def}).

For the shape of  $Q_e^*$ as for the eigenvectors our main idea is to
derive a perturbation theory in powers  of $v_c-v$ (and also $\lambda$ for the eigenvectors) in the region where the non-linear terms cannot be neglected (i.e. in the region where $1- Q_e^*$ is not small)
and to match this expansion with the solution which can be obtained in
the linear region (where $1- Q_e^* \ll 1 $).

\bigskip

\textbf{The shape of $Q_e^*$~:} \\

It is   convenient to consider the solution $Q_v(x)$ of (\ref{eq:defQestar})
\begin{equation}
Q_v''-v Q_v' + g(Q_v)=0
\label{A1}
\end{equation}
on the infinite line such that $Q_v(x) \to 1$ as $x \to - \infty$ and 
 $Q_v(x) \to 0$ as $x \to + \infty$  
(instead of the solution $Q_e^*(x)$ on the semi infinite line).
Here $g(Q)= \sum_k \beta_k (Q^k - Q)$.
The boundary conditions  at $\pm \infty$ determine $Q_v$ up to a
translation.  One particular solution may be selected by imposing that
\begin{equation}
Q_v(x_0)=\frac{1}{2}
\label{A2}
\end{equation}
where $x_0$ is a fixed position (which can be chosen arbitrarily and of course will play no role in our final results).
For $v<v_c= 2 \sqrt{g'(1)}$,  the  solution of (\ref{A1})  has damped
oscillations,  as $x \to -\infty$, of the form
\begin{equation}
Q_v(x)= 1 + U_v \sin \left( \frac{\pi (x + \Phi_v)}{L} \right) \exp \left(\frac{v x }{2} \right) + O\left( e^{v x }\right)
\label{A3}
\end{equation}
where the length $L$ is defined by~:
\begin{equation}
 L = 2 \pi  (v_c^2 - v^2)^{-\frac{1}{2}}
\label{A4}
\end{equation}
so that for large $L$, one gets
\begin{equation}
v \simeq v_c - \frac{2 \pi^2}{ v_c L^2}
\label{A4a}
\end{equation}
The constants $U_v$ and $\Phi_v$   in (\ref{A3}) are a priori complicated functions of $v$ (or $L$) and also depend on $x_0$.

If  $y_L$ is  the position of the right-most zero of $1-Q_v(x)$, then the
solution $Q_e^*$ of (\ref{eq:defQestar}) we are looking for is given by~:
\begin{equation}
Q_e^*(x)=Q_v(x+y_L)
\label{A4b}
\end{equation}
As $v$ approaches $v_c$, the length $L$ diverges, and  (\ref{A3}) allows
one to estimate $y_L$ up to corrections exponentially small in $L$
\begin{equation}
y_L \simeq - L - \Phi_v  .
\label{A4c}
\end{equation}

We now assume that the solution $Q_{v_c}$  of (\ref{A1}) is known for $v=v_c$. One 
can then expand $Q_v(x)$ in powers of $\frac{1}{L}$ or
in powers of $v_c-v$ 
\begin{equation}
Q_v = Q_{v_c} + {1 \over L^2} R + .... = 
 Q_{v_c} + {(v_c -v )}  {v_c \over 2 \pi^2} R + .... 
\label{A4d}
\end{equation}
Putting (\ref{A4d}) into (\ref{A1}) one gets that $R$ should satisfy
\begin{equation}
R''-v_c R' + g'(Q_{v_c}) R = - {2 \pi^2 \over v_c} Q_{v_c}' .
\label{A5}
\end{equation}
One can solve this equation with the required boundary conditions  (the equations leading to higher order terms in the expansion could be solved as well) and one gets~:
\begin{equation}
 R(x) = {2 \pi^2 \over v_c}  Q_{v_c}' (x)  \int_{x_c}^x \frac{dy}{ [
Q_{v_c}'(y)]^2 e^{-vy} }\int_y^{+\infty} [ Q_{v_c}'(z)]^2 e^{-vz} dz.
\label{A6}
\end{equation}
For $x \to - \infty$,
one can show from
(\ref{A1}) that 
\begin{equation}
 Q_{v_c}(x)  =  1 + (A_c x + B_c) e^{ \frac{v_c x}{2} } + H_2(x) e^{v_c x} + H_3(x) e^{\frac{3 v_c x}{2}} + \ldots
\label{A7}
\end{equation}
where the coefficients $A_c$ and $B_c$ depend on $x_0$ and the polynomes $H_n$  could be obtained explicitly in terms of these coefficients $A_c$ and $B_c$ by  analyzing the non-linear terms of (\ref{A1}).

One can also show from (\ref{A6}) 
 or (even better) directly from (\ref{A5}) that for $x \to - \infty$~:
\begin{equation}
R(x) =  \pi^2 \left( - \frac{A_c x^3}{6} - \frac{B_c x^2}{2} -\frac{A_c  x^2}{v_c} + C x + D \right) e^{ \frac{v_c x}{2} }  + O\left(e^{v_c x} \right)
\label{A8}
\end{equation}
where the constants $A_c$ and $B_c$ are the same as in (\ref{A7}) and the constants $C$ and $D$ could be determined from the knowledge of $Q_{v_c}(x)$ and (\ref{A6}).

This leads (\ref{A4d},\ref{A7},\ref{A8}) for large negative $x$ to
\begin{eqnarray}
Q_v(x)&=& 1 + \left[   A_c x + B_c + {\pi^2 \over L^2} \left( - {A_c x^3 \over
6} - {B_c x^2 \over 2} -{\frac{A_c}{v_c} x^2 } + C x + D \right) + .... \right]  e^{ {v_c x \over 2}  } \nonumber \\
& &+ O( e^{v_c x})
\label{A9}
\end{eqnarray}
On the other hand for large negative $x$, in the range $1 \ll |x| \ll L$,
expression (\ref{A3}) becomes
\begin{equation}
Q_v(x)= 1 + U_v \left[
\frac{\pi  (x + \Phi_v)}{L} - \frac{\pi ^3  (x + \Phi_v)^3}{6 L^3} - {
\pi^3 x ( x + \Phi_v) \over v_c L^3} + O \left( L^{-5} \right)
\right] e^{ {v_c x \over 2}  } + ...
\label{A10}
\end{equation}
The comparison of (\ref{A9}) and (\ref{A10}) leads to
the following expansions for $U_v$ and $\Phi_v$~:
\begin{subequations}
\label{A11}
\begin{eqnarray}
U_v &=& \frac{L}{\pi}  A_c + \frac{ \pi}{L} \left( C + \frac{B_c}{v_c} + \frac{B_c^2}{2 A_c} \right) + O \left( \frac{1}{L^3} \right) \\
\Phi_v &=& \frac{B_c}{A_c }   + { \pi^2 \over L ^2} \left(  \frac{D}{A_c} - \frac{B_c^3}{3 A_c^3 } - \frac{B_c^2}{A_c^2 v_c} - \frac{B_c C}{A_c^2} + O \left( \frac{1}{L^4} \right)
 \right) 
\end{eqnarray}
\end{subequations}
Then using (\ref{A4b},\ref{A4c}) one gets the following expressions for
$Q_e^*(x)$ in the region where $L-x\gg{}1$
\begin{equation}
Q_e^*(x) = 1 - \frac{A_c L}{\pi}  \exp\left[ \frac{v_c  (x - L - \frac{B_c}{A_c})
}{2 }\right]
\left[\sin \left(\frac{\pi x }{L}\right) + O\left( \frac{1}{L^2}\right) \right] + O\left(e^{ v_c (x -
L )} \right)
\end{equation}
 and  in the region $x>L$ or $x-L = O(1)$~:
\begin{equation}
Q_e^*(x) =  Q_{v_c} \left(x-L- \frac{B_c}{A_c} \right) + O\left( \frac{1}{L^2}\right)
\end{equation}
We emphasize that the knowledge of $Q_{v_c}(x)$ is sufficient to determine
all the higher order corrections in $\frac{1}{L}$ expansions.

\bigskip

\textbf{The  eigenvalues and the eigenvectors $\phi_\lambda$}~:\\

We consider now the shape of the eigenvectors $\phi_{\lambda,v}$ of the
linear operator $\Lin$ on the infinite line. On the infinite
line, to each $\lambda$, one can associate an eigenvector $\phi_{\lambda,v}$ which satisfies~:
\begin{equation}
\phi_{\lambda,v}''-v  \phi_{\lambda,v}' + g'(Q_v) \phi_{\lambda,v}=
\lambda \phi_{\lambda,v}
\label{A20}
\end{equation}
with the boundary conditions (\ref{eq:eigenboundary}) $\phi_{\lambda,v} (x) \to 0$ as $ x
\to \pm \infty$.
For $v < v_c$, the  solution when $x \to - \infty$ has the form
\begin{equation}
\phi_{\lambda,v}(x)=  V_{\lambda,v} \sin \left( {\pi (x +
\Psi_{\lambda,v})   \over L_{\lambda,v}} \right) \exp \left( \frac{v x}{2} \right) + O\left( e^{v x } \right)
\label{A21}
\end{equation}
 where the length $L_{\lambda,v}$ is defined by
\begin{equation}
L_{\lambda,v} = 2 \pi 
 (v_c^2 -4 \lambda - v^2)^{-{1 \over 2}} = L \left( 1 - { L^2 \lambda \over
 \pi^2} \right)^{-{1 \over 2}}.
\label{A22}
\end{equation}

On the other hand, it is easy to see  that  $Q_{v_c}'(x)$ is an
eigenvector  for  $\lambda=0$ and $v=v_c$,
 (in fact for any $v$, one
has  $\phi_{0,v}(x)=Q_{v}'(x)$ up to a multiplicative  constant) 
and one can try to expand $\phi_{\lambda,v}$ in powers of $\lambda$
and $v_c-v$
\begin{equation}
 \phi_{\lambda,v}(x) =  Q_{v_c}'(x) + \sum_{n + m \geq 1} S_{nm}(x)
{\lambda^n   \over L^{2 m}}
\end{equation}
The $S_{nm}$'s could be  determined recursively solving inhomogenous equations similar
to  (\ref{A5}). For example
\begin{eqnarray*}
 S_{10}'' -v_c S_{10}' + g'(Q_{v_c})  S_{10} &=&    Q_{v_c}' \\
 S_{01}'' -v_c S_{01}' + g'(Q_{v_c}) S_{01} &=& - { 2 \pi^2 \over v_c} Q_{v_c}'' -
g''(Q_{v_c}) R Q_{v_c}'
\end{eqnarray*}
As  in (\ref{A9}) one gets that for large negative $x$
\begin{equation}
\phi_{\lambda,v}(x) = \left[ {A_c v_c \over 2} x + {B_c v_c \over 2} + A_c  +
O(\lambda) + O\left( {1 \over L^2 } \right)  \right]
e^{ {v_c x \over 2}  } + O\left( e^{ {v_c x }  }\right)
\label{A23}
\end{equation}
Comparing  (\ref{A23}) with (\ref{A21})  in the range of large
negative $x$ with $x\ll L_{\lambda,v}$ (as in (\ref{A9},\ref{A10}))
one gets
that
\begin{subequations}
\begin{eqnarray}
\Psi_{\lambda,v} &=& {B_c\over A_c} + {2 \over v_c} + O \left({1 \over L^2}\right) + 
O\left({1 \over L_{\lambda,v}^2}\right)
\label{A24} \\
V_{\lambda,v} &=& {A_c v_c \over 2 \pi} L_{\lambda,v} + O \left({1 \over L}\right) +
O\left({1 \over L_{\lambda,v}}\right)
\end{eqnarray}
\end{subequations}

On the infinite line, as $\lambda$ varies, all the eigenvectors $\phi_{\lambda,v}$ satisfy the
boundary conditions at $x\to \pm \infty$. In the presence of a wall,  the
eigenvector $\phi_{\lambda,v}$  has to vanish  at the wall,  therefore the rightmost
zero  $x=-L-\Phi_v$ of $Q_v(x)$ must coincide with a zero $x=-nL_{\lambda,v}-\Psi_{\lambda,v}$ of $\phi_{\lambda,v}$. The
boundary conditions select that way a discrete set of eigenvalues
$\lambda_n$. For
$v_c-v$ small, one gets then, up to exponentially small corrections in $L$~:
\begin{equation}
\label{eq:zerosloc}
 n L_{\lambda,v}- L = \Phi_v- \Psi_{\lambda,v} 
\end{equation}
where $n$ is an integer larger or equal to  $1$.
Using the  expressions (\ref{A11},\ref{A24}) and  (\ref{A22})
this becomes
$$n \left(1 - \frac{L^2 \lambda}{\pi^2} \right)^{- \frac{1}{2}} -1 = -\frac{2}{v_c L} + O\left({1 \over L^3} \right) $$
and this leads to 
\begin{equation}
 \lambda_n \simeq - \frac{ (n^2-1)\pi^2 }{L^2} - \frac{ 4 n^2 \pi^2   }{ v_c L^3}
- \frac{ 12 n^2 \pi^2 }{ v_c^2 L^4}  + O \left( \frac{1}{L^5} \right)
\end{equation}
One should notice that, for $n=1$, $\lambda_1 \sim L^{-3}$ whereas all the other
eigenvalues $\lambda_n$ scale as $L^{-2}$.

In the frame of the wall, the $n$-th eigenvector $\phi_n$ is, up to exponentially small corrections, given by~:
$$\phi_n(x)\simeq  \phi_{\lambda_n,v}( x + y_L)$$
and therefore
in the range where $L-x \ll 1$
one has
$$\phi_n(x) \simeq \frac{A_c v_c  L (-1)^{n-1}}{2  n \pi}  \left[ \sin \left( \frac{ n \pi x}{L}
\right) + O \left( \frac{1}{L^2 } \right) \right] e^{{v_c(x-L) \over 2} -
\frac{v_c B_c}{2 A_c}} + O\left( e^{v_c(x-L)}\right) $$
and in the range where $x-L$ is of order $1$
$$\phi_n(x) \simeq Q'_{v_c} \left( x - L - \frac{B_c}{A_c}\right) .$$

\label{appendix:eigenvalues}



\end{document}